\documentclass[iop]{emulateapj}		

\usepackage{apjfonts, natbib}
\usepackage{graphicx}

\newcommand{\tnm}[1]{\tablenotemark{#1}}	
\newcommand{\tnt}[1]{\tablenotetext{#1}}	

\newcommand{\av}{\mbox{$A_V$}}			
\newcommand{\avmw}{\mbox{$A_{V,\,{\rm MW}}$}}	
\newcommand{\avint}{\mbox{$A_{V,\,{\rm i}}$}}	

\newcommand{\vk}{\mbox{$V\!-\!K$}}		

\newcommand{\czsyshel}{\mbox{$cz_{\rm sys,hel}$}}
\newcommand{\czlg}{\mbox{$cz_{_{\rm LG}}$}}	
\newcommand{\dvlos}{\mbox{$\Delta V_{_{\rm LOS}}$}}
\newcommand{\ergcms}{\mbox{{\rm erg cm}$^{-2}$ {\rm s}$^{-1}$}}
\newcommand{\ergsec}{\mbox{{\rm erg s}$^{-1}$}}	
\newcommand{\ergsechz}{\mbox{{\rm erg s}$^{-1}$ {\rm Hz}$^{-1}$}}
\newcommand{\etal}{et al.}			
\newcommand{\fzeromw}{\mbox{$F_{0,{\rm MW}}$}}  
\newcommand{\hi}{\ion{H}{1}}			
\newcommand{\hii}{\ion{H}{2}}			
\newcommand{\hiisone}{\mbox{\ion{H}{2}\,(S101)}}
\newcommand{\hst}{{\em HST}}			
\newcommand{\kms}{\mbox{km~s$^{-1}$}}		
\newcommand{\lapp}{\mbox{$L_{\rm ap}$}}         
\newcommand{\lappdisk}{\mbox{$L_{\rm ap,\,D}$}}	
\newcommand{\lappneb}{\mbox{$L_{\rm ap,\,neb}$}} 
\newcommand{\lappnuc}{\mbox{$L_{\rm ap,\,nuc}$}} 

\newcommand{\lineratzero}{\mbox{$F_0/F_0({\rm H}\beta)$}}
\newcommand{\lineratzeromw}{\mbox{$F_{0,{\rm MW}}/F_{0,{\rm MW}}({\rm H}\beta)$}}
\newcommand{\lion}{\mbox{$L_{\rm ion}$}}	
\newcommand{\lionmin}{\mbox{$L_{\rm ion,min}$}}	
\newcommand{\lradio}{\mbox{$L_{\rm 1.4\,GHz}$}}	
\newcommand{\lradionuc}{\mbox{$L_{\rm 1.4\,GHz,\,nuc}$}}
\newcommand{\lxzero}{\mbox{$L_{\rm X,\,0}$}}	
\newcommand{\lxttzero}{\mbox{$L_{\rm 2-10\,\,keV,\,0}$}}
\newcommand{\msun}{\mbox{$M_{\odot}$}}		
\newcommand{\msunyr}{\mbox{$M_{\odot}$ yr$^{-1}$}}

\newcommand{\n}{NGC~}				
\newcommand{\nel}{\mbox{$N_{\rm e}$}}		
\newcommand{\reccohbeteff}{\mbox{$\alpha_{{\rm H}\beta}^{\rm eff}$}}
\newcommand{\reccoHtwo}{\mbox{$\alpha_{\rm B}({\rm H}^{+},T)$}}
\newcommand{\reccoHB}{\mbox{$\alpha_{\rm B}$}}

\newcommand{\sigvelgas}{\mbox{$\sigma_{v,{\rm g}}$}}
\newcommand{\trec}{\mbox{$\tau_{\rm rec}$}}	
\newcommand{\vrot}{\mbox{$v_{\rm rot}$}}	
\newcommand{\vshock}{\mbox{$v_{\rm sh}$}}	
\newcommand{\zsun}{\mbox{$Z_{\odot}$}}		
\newcommand{\halp}{\mbox{${\rm H}\alpha$}}	
\newcommand{\hbet}{\mbox{${\rm H}\beta$}}	
\newcommand{\hgam}{\mbox{${\rm H}\gamma$}}	
\newcommand{\hdel}{\mbox{${\rm H}\delta$}}	
\newcommand{\hei}{\mbox{\ion{He}{1}}}		
\newcommand{\heisix}{\mbox{\ion{He}{1} $\lambda$5876}}   
\newcommand{\heii}{\mbox{\ion{He}{2}}}		
\newcommand{\heiisix}{\mbox{\ion{He}{2} $\lambda$4686}}  
\newcommand{\niboth}{\mbox{[\ion{N}{1}] $\lambda\lambda$5198, 5200}}
\newcommand{\nii}{\mbox{[\ion{N}{2}]}}		
\newcommand{\niifiv}{\mbox{[\ion{N}{2}] $\lambda$5755}}
\newcommand{\niiboth}{\mbox{[\ion{N}{2}] $\lambda\lambda$6548, 6583}}
\newcommand{\niieig}{\mbox{[\ion{N}{2}] $\lambda$6548}}
\newcommand{\niithr}{\mbox{[\ion{N}{2}] $\lambda$6583}}
\newcommand{\neiii}{\mbox{[\ion{Ne}{3}]}}	
\newcommand{\neiiieig}{\mbox{[\ion{Ne}{3}] $\lambda$3868}}
\newcommand{\nev}{\mbox{[\ion{Ne}{5}]}}		
\newcommand{\nevfou}{\mbox{[\ion{Ne}{5}] $\lambda$3426}}
\newcommand{\nevboth}{\mbox{[\ion{Ne}{5}] $\lambda\lambda$3346, 3426}}
\newcommand{\oi}{\mbox{[\ion{O}{1}]}}		
\newcommand{\oizer}{\mbox{[\ion{O}{1}] $\lambda$6300}}
\newcommand{\oii}{\mbox{[\ion{O}{2}]}}		
\newcommand{\oiiboth}{\mbox{[\ion{O}{2}] $\lambda\lambda$3726, 3729}}
\newcommand{\oiisev}{\mbox{[\ion{O}{2}] $\lambda$3727}}
\newcommand{\oiii}{\mbox{[\ion{O}{3}]}}		
\newcommand{\oiiithr}{\mbox{[\ion{O}{3}] $\lambda$4363}}

\newcommand{\oiiisev}{\mbox{[\ion{O}{3}] $\lambda$5007}}
\newcommand{\oiiiboth}{\mbox{[\ion{O}{3}] $\lambda\lambda$4959, 5007}}
\newcommand{\oiiineb}{\mbox{[\ion{O}{3}] nebula}}
\newcommand{\sii}{\mbox{[\ion{S}{2}]}}		
\newcommand{\siiboth}{\mbox{[\ion{S}{2}] $\lambda\lambda$6716, 6731}}

\slugcomment{Submitted to ApJ\, 2013 April 25; accepted\, 2013 July 02}

\shorttitle{The \oiii\ Nebula of \n7252}
\shortauthors{Schweizer et al.}

\begin{document}

\title{The \oiii\ Nebula of the Merger Remnant \n7252: A Likely \\
       Faint Ionization Echo\altaffilmark{1}}

\author{
Fran\c cois Schweizer\altaffilmark{2},
Patrick Seitzer\altaffilmark{3},
Daniel D.\ Kelson\altaffilmark{2},
Edward V.\ Villanueva\altaffilmark{2}, \\
and Gregory L.\ Walth\altaffilmark{4}
}

\altaffiltext{1}{Based in part on data gathered with the 6.5 m Magellan
Telescopes located at Las Campanas Observatory, Chile.}
\altaffiltext{2}{Carnegie Observatories, 813 Santa Barbara Street, Pasadena,
   CA 91101; schweizer@obs.carnegiescience.edu}
\altaffiltext{3}{Department of Astronomy, University of Michigan, 818 Dennison
   Building, Ann Arbor, MI 48109}
\altaffiltext{4}{Steward Observatory, University of Arizona, 933 N. Cherry Ave,
   Tucson, AZ 85721}



\begin{abstract}

We present images and spectra of a $\sim\,$10~kpc-sized emission-line
nebulosity discovered in the prototypical merger remnant \n7252 and dubbed
the `\oiiineb' because of its dominant \oiiisev\ line.
This nebula seems to yield the first sign of episodic AGN activity still
occurring in the remnant, $\sim\,$220~Myr after the coalescence of two
gas-rich galaxies.
Its location and kinematics suggest it belongs to a stream of tidal-tail
gas falling back into the remnant.
Its integrated \oiiisev\ luminosity is $1.4\times10^{40}$ \ergsec, and
its spectrum features some high-excitation lines, including \heiisix.
In diagnostic line-ratio diagrams, the nebula lies in the domain of Seyfert
galaxies, suggesting that it is photoionized by a source with a power-law
spectrum.
Yet, a search for AGN activity in \n7252 from X-rays to radio wavelengths
yields no detection, with the most stringent upper limit set by X-ray
observations.
The upper luminosity limit of $\lxttzero<5\times10^{39}$ \ergsec\ estimated
for the nucleus is $\sim\,$10$^3$ times lower than the minimum ionizing
luminosity of $\ga5\times10^{42}$ \ergsec\ necessary to excite the nebula.
This large discrepancy suggests that the nebula is a faint ionization echo
excited by a mildly active nucleus that has declined by $\sim\,$3 orders of
magnitude over the past 20,000\,--\,200,000 years.
In many ways this nebula resembles the prototypical `Hanny's Voorwerp'
near IC~2497, but its size is 3$\times$ smaller and its \oiii\ luminosity
$\sim\,$100$\times$ lower.
We propose that it be classified as an extended emission-line region (EELR).
The \oiiineb\ is then the lowest-luminosity ionization echo and EELR
discovered so far, indicative of recent, probably sputtering AGN activity
of Seyfert-like intensity in \n7252.

\end{abstract}



\section{INTRODUCTION}
\label{sec1}

Over recent decades the hierarchical assembly of galaxies \citep{white78} has
become a paradigm of modern $\Lambda$ cold dark matter cosmology.
Whereas the assembly of dark-matter halos is now reasonably well modeled and
understood \citep[e.g.,][]{fw12}, the {\em baryonic} growth of galaxies
through gas accretion \citep{rees77,katz91,kere05,deke06} and mergers
\citep{tt72,bh92,bh96,mh96,hopk06} has turned out to be much more challenging
to simulate.
Some key issues of current theoretical and observational interest are the
relative contributions of gas accretion and mergers to galaxy assembly, the
apparent coevolution of supermassive black holes and spheroids 
\citep{mago98,gebh00,ferr00}, and the effects of gaseous outflows in
slowing---or possibly even quenching---the galactic growth process
\citep[e.g.,][]{hopk06,hopk13}.

Mergers of gas-rich galaxies can offer insights on at least two of these
issues.
Especially major mergers (i.e., those with mass ratios $m/M\ga 1/3$) often
lead to intense starbursts and even some AGN activity, each of which can
drive strong gaseous outflows \cite[e.g.,][]{heck90,veil05,stur11}.
In trying to assess the relative influence of starbursts and AGN activity
on the subsequent evolution of merger remnants, their individual durations
and relative intensities matter, as does the detailed sequence of their
occurrence.
Whereas starbursts tend to peak close to the time of coalescence of the
nuclei \citep[e.g.,][]{sand96}, AGN activity may last longer and be more
sporadic.
Observational studies of gas-rich mergers and their remnants have found
peak AGN activity delayed by typically $\sim\,$10$^8$--10$^9$ yr relative
to the peaking of starbursts \citep{cana01,kauf03,scha07,wild10}, in rough
agreement with theoretical estimates based on $N$-body plus hydrodynamical
simulations suggesting AGN delays of 10$^7$ yr to a few 10$^8$ yr, with a
wide scatter \citep{hopk12}.

A problem for all studies involving not only ongoing mergers, but also
merger remnants is the uncertainty about the remnants' past merger
history: What kinds of galaxies merged, and what were their gas contents
and mass ratios?
Our ignorance of such important details in well-known active merger remnants
as, e.g., \n5128 and \n1316 is both surprising and embarrassing.
Clearly, a few cases of well-studied merger remnants with known merger
history could provide welcome checks on observational claims based on
larger, statistical studies.

In this regard, the prototypical merger remnant \n7252 would seem to hold
great promise.
It is the only {\em remnant\,} whose past merger history has been simulated
in some detail \citep{born91,miho93,himi95,miho98,chba10}.
There is little doubt that two gas-rich spirals began merging less than
1~Gyr ago, coalesced about 220 Myr ago, and have transformed themselves
into some sort of modern-day protoelliptical \citep{t77,s82,schw98}.
Much is known about its gas contents and kinematics \citep{wang92,hibb94},
post-starburst stellar populations \citep{s82,schw90,liu95}, and system of
massive young star clusters formed during the merger
\citep{whit93,mill97,ss98,bc03,bast13}.

Yet, until the present study no sign of AGN activity in \n7252 was found,
leaving us in the dark concerning the possible presence of any supermassive
black hole (SMBH), its mass, activity, and recent growth history.
Admittedly, one could construe the presence of considerable amounts of
molecular gas in the small central disk
\citep[$\sim$\,$5\times 10^9\,\msun$:\,][]{dupr90,wang92}
and of neutral hydrogen in the tidal tails
\citep[$3.8\times 10^9\,\msun$:\,][]{hibb94}
as evidence that \n7252 experienced no quasar phase in its recent history.
Yet, gas from the tidal tails keeps falling back into the remnant
\citep{hibb94,himi95,chba10} and replenishing the central gas disk.
Hence, it may be obliterating earlier evidence---if it existed---that strong
AGN activity expelled most or all of the central gas.

Our serendipitous discovery of an extended emission-line nebulosity in
\n7252 promises to shed new light on the question of the presence of a
central SMBH and its past and present activity.
We discovered the existence of this faint nebulosity in 1991 August during
a cloudy observing run with the Modular Spectrograph at the du Pont 2.5 m
telescope on Las Campanas.
Hoping to exploit occasional gaps in the drifting cirrus, we took several
hour-long exposures with the spectrograph slit placed across the nucleus of
\n7252 at position angles corresponding to the major and minor axes of the
central ionized-gas disk \citep{s82}.
A lone, faint and extended emission line struck us as the only spectral
feature visible {\em beyond\,} this central gas disk; it appeared about
10\arcsec--\,20\arcsec\ SSW of the nucleus along the projected rotation
axis of the disk and was quickly identified as \oiiisev\ emission at the
redshift of the galaxy.
Hoping to perhaps find some high-excitation outflow or major shock front,
we followed up in 2000 September by imaging \n7252 through narrow-band
\oiiisev\ and \halp\ filters, also with the du Pont telescope.
The resulting images, presented in \S~\ref{sec21} below, revealed the
presence of a $\sim$\,10 kpc large nebulosity significantly brighter in
\oiiisev\ than in \halp, whence we dubbed it the `\oiiineb.'
It is this nebula that is the central subject of the present paper.

In the following, Section~\ref{sec2} describes our observations and
reductions, including broad- and narrow-band imaging and long-slit
spectroscopy of the nebula.
Section~\ref{sec3} presents results concerning the general morphology of the
ionized gas in \n7252 and the properties of the \oiiineb, including its
structure, metallicity, spectrum, excitation, and kinematics.
Section~\ref{sec4} then discusses excitation mechanisms for the nebula,
some properties of the nucleus of the galaxy, the hypothesis that the
nebula is an ionization echo, its properties compared to those of other
known such echos, and some implications for the evolutionary history of
\n7252.
Finally, Section~\ref{sec5} summarizes our main results and conclusions.

\section{OBSERVATIONS AND REDUCTIONS}
\label{sec2}

For general reference, the merger remnant \n7252 is also known as Arp 226
and the ``Atoms for Peace" galaxy.
It is located at $\alpha_{\rm J2000}=22^{\rm h}20^{\rm m}44\fs78$,
$\delta_{{\rm J}2000}=-24\degr40\arcmin41\farcs8$ \citep{mill97}
and has a recession velocity relative to the Local Group of
$\czlg = +4831$ \kms\ \citep[value revised by $+$3 \kms]{s82}, which places it
at a distance of 66.2 Mpc for $H_0 = 73$ \kms\ Mpc$^{-1}$ \citep{free10}.
At that distance, adopted throughout the present paper, $1\arcsec = 321$ pc.
The corresponding true distance modulus is $(m-M)_0 = 34.10$.
The Milky Way foreground extinction is small, with values in the literature
ranging between $\avmw = 0.038$ (\citealt{rc3}) and 0.100 \citep{schl98}.
We adopt the latter value, with which the absolute visual magnitude of
\n7252 becomes $M_V = -22.08$.

Our observations of \n7252 described below include a series of direct images
obtained with the Ir\'en\'ee du Pont 2.5-m telescope at Las Campanas
Observatory and spectra taken with two spectrographs on the Clay 6.5-m
telescope there.
Table~\ref{tab01} presents a log of these observations.

\subsection{Imaging}
\label{sec21}

Direct broad- and narrow-band images of \n7252 were obtained with the CCD
camera of the du Pont telescope during the four nights of 2000 September
27--30 (Table~\ref{tab01}).
Conditions were photometric, and the seeing was in the range
$0\farcs6$--$1\farcs0$ (FWHM).

The two broad-band images were taken through standard $BV$ filters.
The camera was equipped with the chip Tek~5 (format 2048$\,\times\,$2048),
which yielded a scale of $0\farcs2602$ pixel$^{-1}$ in $V$ and a field of
view of $8\farcm8\times 8\farcm8$ in $B$ and $V$.

The narrow-band images were taken through three interference filters
centered on the emission lines \oiiisev, \halp, and \siiboth\ at the
systemic redshift of \n7252.
To permit subtraction of the stellar continuum from the galaxy, we also
obtained exposures through two intermediate-band filters named "BluCont"
and "RedCont," as detailed in Table~\ref{tab01}.
The filter pairs (\oiii, BluCont) and (\halp, RedCont) were specially
designed for this study, and the filters were fabricated in $2\times 2$-inch
size by Barr Associates, Inc.
The unvignetted field of view through these four filters was approximately
$7\farcm0\times 7\farcm0$.
The \sii\ filter was borrowed from a filter set belonging to Cerro Tololo
Inter-American Observatory and was used in combination with our
RedCont filter.

The various direct CCD frames were flatfielded, coadded, and reduced in
standard manner with
IRAF.\footnote{
The Image Reduction and Analysis Facility (IRAF) is distributed by the
National Optical Astronomy Observatories (NOAO), which are operated by
the Association of Universities for Research in Astronomy (AURA), Inc.,
under a cooperative agreement with the National Science Foundation.}
The $BV$ images were calibrated through observations of standard
stars in the fields L98 and L110 \citep{land73,land92} observed during
the same nights, while no special calibration frames were taken for
the narrow-band images.
To isolate the emission-line fluxes in the three narrow-band images
from the galaxy continuum, scaled versions of the BluCont and RedCont
images were subtracted, with the scale factors determined by
trial and error to minimize the galaxy background contribution.
The resulting deep \oiiisev\ and \halp\ emission-line images are shown
and discussed below, while the less deep \sii\ emission-line image
featured little measurable flux and is not further discussed here.

\begin{figure*}
  \begin{center}
    \includegraphics[scale=0.77]{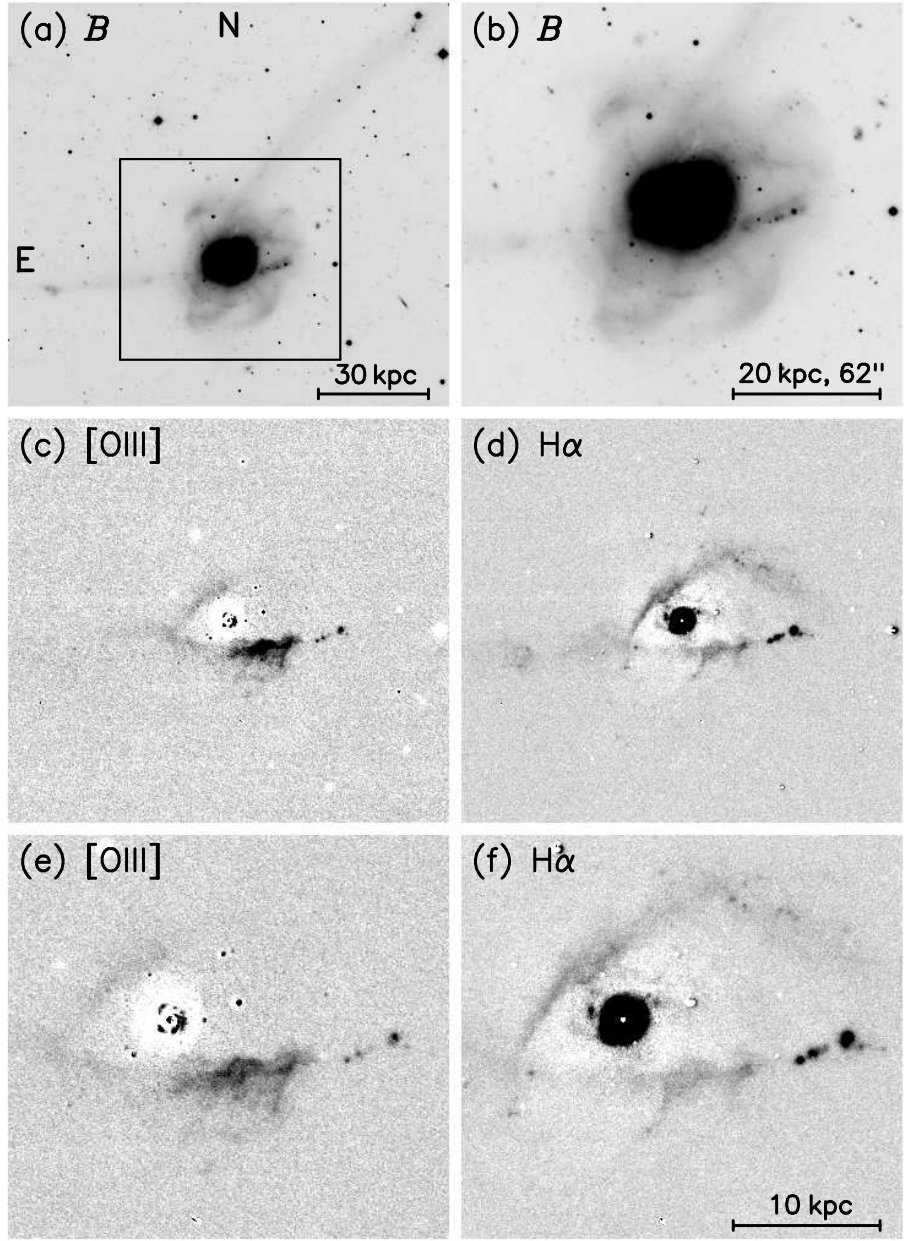}
    \caption{
Groundbased $B$ and narrow-band \oiii\ and \halp\ images of \n7252 obtained
with the du Pont 2.5-m telescope.
({\em a}) High-contrast display of $B$ image, showing $6\farcm2\times5\farcm7$
($120\times 110$~kpc$^2$) field of view; the box marks the $3\farcm1\times
2\farcm85$ ($60\times 55$~kpc$^2$) area displayed in panels (b)\,--\,(d).
Notice the two tidal tails connecting to loops around the main body.
({\em b}) Portion of $B$ image displayed at twice enlarged scale relative
to Panel~(a).
({\em c}) and ({\em d}) Continuum-subtracted \oiiisev\ and \halp\
emission-line images, reproduced at the same scale as Panel (b) and showing
the distribution of ionized gas.
Notice the \oiiineb\ SW of the nucleus in Panel (c).
({\em e}) and ({\em f}) Same \oiii\ and \halp\ images as in panels (c) and
(d), but twice enlarged and reproduced at lower contrast.
Details of the \oiiineb\ and of five \hii\ regions in the W loop are better
visible.
Notice also the brightly \halp-emitting disk centered on the nucleus (black
circular area).
This ionized- and molecular-gas disk corresponds to the central ``minispiral''
observed on \hst\ images;
for details, see text.
    \label{fig01}}
  \end{center}
\end{figure*}

Figure~\ref{fig01} shows portions of the $B$ image (panels a, b),
\oiiisev\ image (panels c, e), and \halp\ image (panels d, f) at
various contrasts, with scale bars indicating the angular and projected
linear scales.
The box in Figure~\ref{fig01}{\em a} identifies the area shown 2$\times$
enlarged in Figures~\ref{fig01}{\em b}--\ref{fig01}{\em d}.
The bottom two panels are enlarged 2$\times$ more, showing a region half
the size (and one quarter the area) of the box drawn in
Figure~\ref{fig01}{\em a}.
Both the \oiiisev\ and the \halp\ images had the galaxy continuum
subtracted as described above.
They are displayed at relatively high contrast in the middle two panels,
and at lower contrast in the bottom two panels to show more details.
Notice the striking \oiii\ emission nebulosity slightly below the
centers of panels (c) and (e), the bright central disk of \halp\ emission
surrounding the nucleus of the galaxy (circular black areas in panels d
and f), and the string of five \hii\ regions $\sim$\,14~kpc west of the
nucleus, best visible in the same two \halp\ panels.

The \oiii\ nebulosity (hereafter `\oiiineb') covers a projected area of
about $10.1\times 6.6$~kpc$^2$ ($31\farcs5\times 20\farcs6$) in the
east--west and north--south directions, respectively, and appears
brightest at a point lying about 4.0~kpc ($12\farcs4$) west and 3.6~kpc
($11\farcs2$) south of the nucleus.
The apparent structure of this \oiiineb\ is discussed in more detail
in \S~\ref{sec31} below.

\subsection{Spectroscopy}
\label{sec22}

Long-slit spectra of the \oiiineb\ in \n7252 were obtained with both 
the Low-Dispersion Survey Spectrograph (LDSS-3; see \citealt{alli94}
for LDSS-2) and the Magellan Echellette (MagE) spectrograph \citep{marsh08}
at the Clay 6.5-m telescope (see Table~\ref{tab01}).
All spectra were taken with the spectrograph slits placed at a position angle
of P.A.\,= $103\fdg3$ across the \oiiineb\ (e.g., Fig.~\ref{fig02}{\em a}
below).

\subsubsection{LDSS-3 Spectra}
\label{sec221}

During the 2007 September observations with LDSS-3, three long-slit
spectra of the nebula were obtained.
The first was taken with the ``VPH Blue" grism (1090 g mm$^{-1}$) through
a $1\farcs1\times 70\arcsec$ slit, yielding a wavelength coverage of
3850--6590~\AA\ and a spectral resolution of 4.0~\AA\ after full reduction.
The second spectrum was taken with the ``VPH Red" grism (660 g mm$^{-1}$)
through the same slit, yielding a wavelength coverage of
6070--10,760~\AA\ and a resolution of 6.6~\AA.
Finally, the third spectrum was taken with the same ``VPH Blue'' grism
as above, but through a mask with four parallel slits of
$0\farcs9\times 70\arcsec$ each and separated by $3\farcs60$ (center to
center) from each other; its spectral resolution is 3.5~\AA.
The purpose of this (third) spectrum was to map ionized-gas velocities
measured from the \oiiisev\ emission line across the nebula, as described
in \S~\ref{sec34}.
The spatial scale along the slit for all three 2D spectra is
$0\farcs1890$ pixel$^{-1}$, with the spatial resolution set by the
average seeing during the observations (Table~\ref{tab01}).

\begin{figure}
  \begin{center}
    \includegraphics[width=8.2cm]{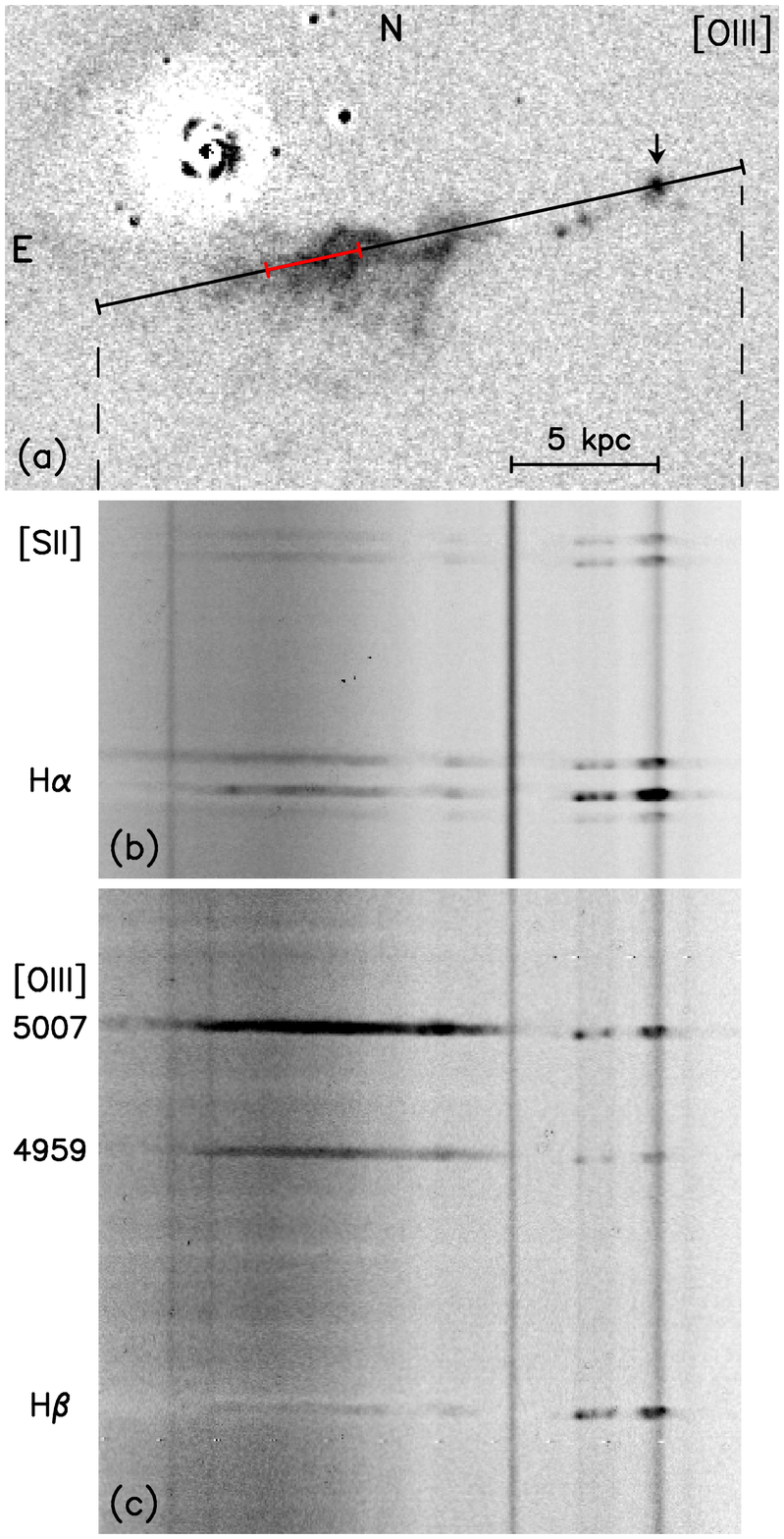}
    \caption{
Sky-subtracted spectra of \n7252's \oiiineb\ obtained with the LDSS-3
spectrograph.
({\em a}) Direct \oiiisev\ image with {\em long black line} showing the
position of the $1\farcs1\times 70\arcsec$ slit placed across the \oiiineb\
and the giant \hii\ region \hiisone\ (marked by downward-pointing arrow);
north is up and east to the left.
({\em b}) Segment of 2D red spectrum showing the nebula's \halp\ emission
line flanked by the \niiboth\ lines and, farther up, the \siiboth\ doublet.
The spectrum is reproduced with its spatial direction (along slit) oriented
horizontally and wavelengths increasing vertically.
({\em c}) Segment of 2D blue spectrum showing \hbet\ and the \oiiiboth\
doublet.
Both spectra cover 70\arcsec\ and are aligned vertically with the projection
of the true slit onto the R.A.\ axis of ({\em a}); they are displayed with
the same intensity scale and zero-point.
The vertical streak seen especially well in the red spectrum stems from a
faint foreground star appearing as a white patch in the continuum-subtracted
\oiii\ image of ({\em a}).
The short red line in ({\em a}) marks the position of the $1\arcsec\times
10\arcsec$ slit of the MagE spectrograph during the spectral exposure
discussed in \S~\ref{sec222}.
    \label{fig02}}
  \end{center}
\end{figure}

Figure~\ref{fig02} shows the position of the main LDSS-3 slit as placed
across the \oiiineb\ during the observations (panel {\em a}), and portions
of the fluxed and sky-subtracted 2D red and blue spectra (panels {\em b}
and {\em c}, respectively).
The $1\farcs1\times 70\arcsec$ slit was designed and positioned to cover
the brightest area of the \oiiineb\ (\S~\ref{sec21}) and to also cut across
the brightest \hii\ region in the western loop of tidal material.
This giant \hii\ region (marked by a downward pointing arrow in panel
{\em a}) is centered on the young star cluster S101 \citep{ss98}, which
served as a {\em spatial} reference point for our various measurements
along the slit; henceforth, it will be referred to as \hiisone.
On the two spectral segments shown in panels {\em b} and {\em c}, notice
the dominance of the \oiiisev\ emission line in the nebula, while
in \hiisone\ \halp\ is clearly the strongest line.
Notice also the broad \hbet\ absorption line underlying the weak \hbet\
emission of the nebula.
This absorption component stems from the galactic background light,
which in \n7252 features a post-starburst spectrum well into the outer
parts of the remnant \citep{s82,schw90,liu95}.

To permit flux calibration, three standard stars (EG~274, Feige~110, and
LTT~1020) were observed repeatedly throughout the run, with both the blue
and red grisms and using a $1\farcs5$ wide slit at parallactic angle.

The subsequent reduction of the various spectra included pipeline
processing to flat-field, wavelength calibrate, and rectify the spectra
frame by frame \citep{kels98,kels00}.
Frames belonging to multiple exposures were then coadded and cleaned of
cosmic rays with the IRAF task {\em imcombine}.
Finally, the resulting blue and red rectified 2D spectra were
flux-calibrated and sky-subtracted in preparation for the various object
extractions and measurements to follow (Fig.~\ref{fig02}{\em c},\,{\em b}).
A careful comparison of the galaxy continuum in the overlap region
($\lambda = 6080$\,--\,6580\,\AA) of the blue and red 2D spectra
showed that, to get equal fluxes there, blue fluxes needed to be multiplied
by, and red fluxes divided by, a factor of 1.027.
These small flux-scale corrections were applied to all fluxes from
LDSS-3 spectra displayed or quoted in the present paper.

\subsubsection{MagE Spectrum}
\label{sec222}

During our 2009 August 23 observations with the MagE spectrograph, the
$1\farcs0\times 10\arcsec$ slit was placed across the brightest part of the
\oiiineb, as indicated by the red bar in Figure~\ref{fig02}{\em a}.
The main goals of taking this spectrum were
(1) to better cover the UV part of the nebular spectrum, including the
\oiisev\ doublet not covered by the LDSS-3 spectra, and
(2) to permit a reliable measurement of the Balmer emission-line decrement.
The 2.5 hr total exposure was broken into five 30 min subexposures,
during which the airmass increased from 1.005 to 1.284 and the seeing
held steady at $0\farcs7$--$0\farcs75$.
A separate sky exposure, also of 30 min duration, was then obtained
with the slit offset to a patch of blank sky 90\arcsec\ N of the
\oiiineb.
To calibrate fluxes, six standard stars were observed with the same
slit at parallactic angle throughout the night.

The subsequent reduction of the various MagE spectra included pipeline
processing to flat-field and coadd frames, rectify spectral orders,
calibrate wavelengths, and subtract the sky spectrum.
Although MagE covers the wavelength range of 3100~\AA\ -- 1.0~$\mu$m in
orders 20\,--\,6 \citep{marsh08}, the two most ultraviolet orders yielded
no signal and the two most infrared orders could not be reliably processed
because of scattered-light problems.
The final extracted spectra cover the wavelength range 3300\,--\,8250~\AA,
extracted from orders 18\,--\,8, at a spectral resolution of $R \approx 4100$.
The spatial scale along the slit varies from $0\farcs251$ pixel$^{-1}$ in
Order 18 (UV) to $0\farcs285$ pixel$^{-1}$ in Order 8 (near-infrared).

Two kinds of measurements were made from the rectified 2D order spectra.
First, emission-line velocities and profiles were measured along the slit
from selected orders, with results described in \S\S~\ref{sec34} and
\ref{sec331} below.
And second, a full spectrum of the brightest part of the \oiiineb\ was
extracted from orders 18\,--\,8 along a $6\farcs61$ segment of the slit
corresponding to a projected area of $2.12\times 0.32$ kpc$^2$.
This segment was carefully chosen to avoid ionized gas with a strong
velocity gradient near the east end of the slit;
it extends from $4\farcs54$ WNW to $2\farcs07$ ESE of the slit center,
which itself was positioned at
$(\Delta\alpha, \Delta\delta) = (-11\farcs78, -11\farcs95)$
relative to the galaxy nucleus.
For each spectral order, a 1D spectrum was extracted from the described
segment of the slit, taking the order-dependent spatial scale into account,
and the resulting eleven 1D spectra were then spliced into a single spectrum
on a linear wavelength scale via a specially written IRAF script.
In a final step, the weak stellar background spectrum from the outskirts of
the galaxy was extracted, slightly smoothed, and then subtracted to produce
a purely nebular emission-line spectrum.

Figure~\ref{fig04} in \S~\ref{sec33} below shows the resulting
flux-calibrated spectrum of the brightest part of the \oiiineb, plotted vs
rest wavelength and with detected nebular emission lines
identified.\notetoeditor{
Please do not insert Figure 4 here, where it only gets a brief advance
reference.  Instead, Fig.~4 must follow Fig.~3 and remain in Section 3.3,
where it gets discussed in detail.}
This spectrum forms the centerpiece of the present study and is discussed
in detail in \S~\ref{sec33}.

\section{RESULTS}
\label{sec3}

The following subsections first briefly describe the general morphology
of the ionized gas in \n7252 and where the \oiiineb\ fits in, then derive
the metallicity of the brightest giant \hii\ region in the same gas stream
as the nebula, and go on to present the spectral properties, physical
properties, and kinematics of the nebula in detail.

\subsection{Morphology of Ionized Gas in \n7252}
\label{sec31}

Ionized gas pervades the main body of \n7252 and shows complex kinematics
\citep{s82,hibb94}, as a future paper will describe in more detail.
Here we give a brief overview of its apparent morphology to show where the
\oiiineb\ fits in and point out some details of importance for the
interpretation of its spectrum.

As the panels of Figure~\ref{fig01} labeled ``\oiii'' and ``\halp''
show, the projected distribution of the ionized gas can be naturally divided
into four main areas:
(1) a central gas disk very bright in \halp;
(2) a NE stretch (or ``front'') best seen in \halp\ and lying roughly
parallel to the NW tail of tidal debris seen in panels {\em a} and {\em b};
(3) a long southern stretch of gas best visible in \halp\ (panel {\em d}) and
extending from the E tail westward past the central disk and through the
region of the \oiiineb\ all the way to a string of bright \hii\ regions
ending with \hiisone; and
(4) a faint NW stretch of gas seen only in \halp\ and containing
various faint and fuzzy \hii\ regions (panels {\em d} and {\em f}).

The {\em central disk} of ionized and molecular gas extends to a radius of
about 7\arcsec\ (2.2 kpc) and rotates with
$\vrot \sin i \approx 100$\,--\,110 \kms, corresponding to
$\vrot \approx 160 \pm 30$~\kms\ with the uncertainty of the disk
inclination ($i = 41\degr \pm 9\degr$) included \citep{s82,wang92}.
As can be seen on \hst\, broad-band images, it hosts a large number of
young star clusters and OB associations arranged in spirally patterns,
leading to its nickname ``minispiral'' \citep[esp.\ Fig.~8]{whit93,mill97}.
An interesting peculiarity is that the gas disk seems to feature a small
``hole'' at the nucleus, first noticed in the ionized gas as a likely
``1\arcsec\,--\,2\arcsec\ hole near the center'' by \citet{s82} and then
also seen in the molecular gas (CO) distribution by \citet{wang92}.
This hole, if representing a central cavity of a few hundred parsecs
diameter in the gas distribution, may tell us something about recent
activity in the nucleus (see \S~\ref{sec45}).

The {\em NE stretch} of ionized gas, brighter in \halp\ than in \oiiisev\
(Figs.~\ref{fig01}{\em d} and \ref{fig01}{\em c}), coincides with a\ $\sim$15
kpc long dust lane (or system of dust lanes) along the inner edge of the NW
tail, as can be seen by comparing Figures 5b and 5d of \citet{hibb94}.
This extensive dust lane makes it clear that the NW tail passes {\em in front}
of the main body before looping around it in the southeast \citep{s82}.
By association, the NE stretch of ionized gas must lie in front of the main
stellar body of \n7252 as well.

Similarly, the {\em S stretch} of ionized gas is clearly associated with
the stellar debris of the E tail, best seen in Figures~\ref{fig01}{\em a}
and \ref{fig01}{\em b}.
Since the \oiiineb\ (Fig.~\ref{fig01}{\em c}) appears well lined up with the
general \halp\ emission along that tail (Fig.~\ref{fig01}{\em d}), it seems
likely to be part of the tail gas.
In \S~\ref{sec34} we will use its kinematics to verify that this indeed
seems to be the case.
The E tail itself passes {\em behind} the main body of \n7252, as suggested
by the strongly negative line-of-sight velocity ($-$230 \kms) of \hiisone\
near its west end (\S~\ref{sec34}), the kinematics of the \hi\ gas along the
tail \citep{hibb94}, and $N$-body simulations reproducing the observed tail
shapes and velocities \citep{himi95,chba10}.
Hence, the \oiiineb\ itself probably also lies behind the main body of
the remnant.

The faint {\em NW stretch} of ionized gas is more difficult to place in
space.
By its projected location and orientation, it seems to form part of the
``W Loop'' seen on deep photographs \citep{s82} and in
Figure~\ref{fig01}{\em b}.
If so, it may be a continuation of the E tail material that is falling back
into the remnant \citep{himi95}.
The line-of-sight velocity of about $-$60 \kms\ for one of its faint \hii\
regions (unpublished) seems to agree at least in sign, if not necessarily
in magnitude, with this interpretation.
The \hi\ distribution also seems to indicate continuity along the loop
\citep{hibb94}, though with coarse resolution.
Yet, the continuity cannot be established beyond any doubt from our $B$, $V$,
and \halp\ images, which all fail to show a direct luminous connection between
emission from the NW stretch and emission from the E tail that seems to end
just past \hiisone.
Perhaps the apparent gap in luminosity is due to some dust obscuration,
or else the W ``loop'' is only apparent.
We do not pursue this detail any further here, given that it does not directly
impact our discussion of the \oiiineb\ and its properties.

The apparent structure of the \oiiineb\ itself is, of course, of interest.
Perhaps the most striking features are the apparent filaments (``streamers'')
extending from the ridge of highest surface brightness in \oiiisev\ to the
south.
Notice their different relative brightnesses in \oiii\ and \halp.
Whereas the westernmost filament appears brightest in \oiii, it is the
easternmost that appears brightest in \halp.
Another feature of interest is a possible dust lane that {\em may\,} run
along the E-W ridge of highest \oiii\ surface brightness in a step-like
fashion (see Fig.~\ref{fig01}{\em e}).
The reality of this possible dust lane is---however---difficult to establish,
given the relatively low signal-to-noise ratio of the \oiii\ image.

We close this discussion of the ionized-gas morphology with a cautionary
note.
In the absence of high-resolution 2D kinematic information, any specific area
of the projected ionized-gas distribution in this messy merger remnant could
lie anywhere along the line of sight, and its preliminary assignment to a
certain depth level within the remnant might have to be revised in the light
of future new evidence.

\begin{figure}
  \begin{center}
    \includegraphics[width=8.4cm]{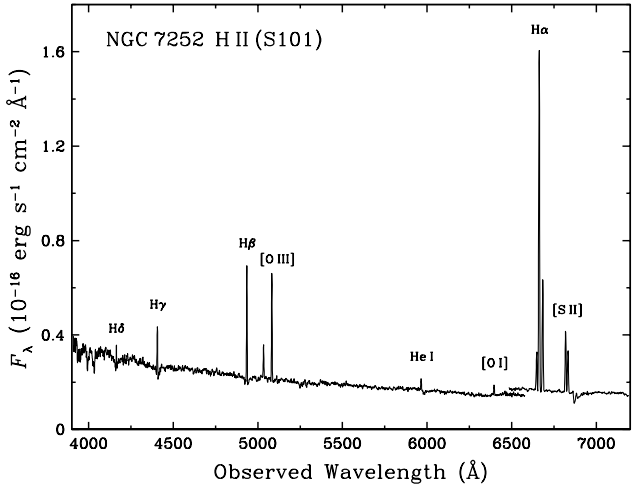}
    \caption{
Blue and red spectra of the giant \hii\ region surrounding Cluster S101,
obtained with LDSS-3 on the Clay 6.5 m telescope.
The main emission lines are labeled (except for the \niiboth\ lines
flanking \halp).
The two spectra have been sky-subtracted and fluxed, but not background
subtracted; for details, see text.
    \label{fig03}}
  \end{center}
\end{figure}

\subsection{Abundances from Giant \hii\ Region Around Cluster S101}
\label{sec32}

In interpreting the unusual spectrum of the \oiiineb, having a reliable,
independent estimate of the metallicity of the involved gas would be helpful.
Fortunately and by design, the main slit used for our spectroscopic
observations of the nebula with LDSS-3 crossed \hiisone, the giant
\hii\ region surrounding Cluster S101 (Fig.~\ref{fig02}{\em a}), yielding a
new spectrum of it.
Therefore, we had a chance to redetermine the metallicity of this \hii\
region, for which \citet{ss98} had measured a logarithmic metallicity relative
to the sun of $[Z/\zsun] = -0.12\pm 0.05$ with the Ritchey-Chr\'etien
spectrograph of the Blanco 4-m telescope.

A blue and a red 1D spectrum of \hiisone\ were extracted from the
corresponding rectified and sky-subtracted 2D spectra (\S~\ref{sec221})
through an aperture measuring $4\farcs72\times 1\farcs1$ (corresponding to
$1.5\times 0.35$ kpc$^2$), where the larger dimension represents the 
extent of detected \halp\ and the smaller dimension the projected slit width;
this aperture was centered on Cluster S101.
Figure~\ref{fig03} displays the two spectra, which show the strong emission
lines typical of a giant \hii\ region, in the observed-wavelength interval
3900\,--\,7200~\AA.
Note that the spectral continuum contains not only light from Cluster S101,
but also some background light from the galaxy.
Disentangling these two continuum contributions proved too difficult, mainly
because of the lack of an atmospheric-dispersion corrector (ADC) at the
Clay telescope.
As a result and despite certain precautions, the spectrum of Cluster
S101---which appears nearly as a point source---experienced different slit
losses as a function of wavelength than did the spectrum of the galaxy
background.
The same cause likely explains the mismatch of the cluster continuum where
the blue and red spectra overlap.
However, the emission lines from \hiisone\ should be affected significantly
less than the cluster continuum since this giant \hii\ region extends over
$\sim$4\arcsec.

Table~\ref{tab02} gives integrated emission-line fluxes measured from the blue and
red spectra via the IRAF task {\em splot}.
Because of the continuum-subtraction problem mentioned above, the spectra
were displayed as shown in Fig.~\ref{fig03} and the ``continuum'' level was
chosen visually near each emission line as best possible.
For the Balmer lines, this involved estimating the profile of the broad
underlying absorption component, whence especially the \hdel\,--\,\hbet\ line
fluxes may be affected by that component more than the pure measuring errors
indicate. 

The mean visual extinction \av\ for the emission-line spectrum of \hiisone\ 
can be estimated from the reddening of the Balmer lines by comparison with
a Case B spectrum for a temperature of $T = 10,000$~K \citep{of06} and is
$\av \approx 1.0\pm 0.2$.
This value includes the Milky Way foreground extinction of $\avmw = 0.10$
\citep{schl98} and is derived from the measured flux ratios for $\hdel/\hbet$
and $\halp/\hbet$ (Table~\ref{tab02}).
(The ratio $\hgam/\hbet$ yielded no reasonable answer, $\av\la 0.0$.)
The above mean extinction agrees with a re-evaluation of the value of
$\av\approx 1.3$ quoted by \citet{ss98} for a Case A spectrum, which for the
more appropriate assumption of Case B emission yields $\av\approx 1.0$ as
well.

Concerning the mean nebular temperature $T$ of \hiisone, the approximate
upper flux limit for the \oiiithr\ line---unusually large because of a likely
noise spike at the observed wavelength---yields only a nominal upper limit
of $T\la 23,000$~K (via the \oiii\ \{4959+5007\}/4363 line ratio), hardly
of interest given the limited temperature range of about 7,000\,--\,14,000~K
for known giant \hii\ regions \citep{of06}.
The observations by \citet{ss98} yielded a stricter upper flux limit for
\oiiithr\ and, from it, $T < 15,000$~K.
In agreement with this upper limit, the abundance-determination method about
to be described yields $T \approx 8450\pm 100$~K.

Chemical abundances for oxygen and nitrogen in \hiisone\ can be
estimated---despite the lack of a measured flux for the \oiiboth\
doublet---through the method developed by \citet{pily11} specifically for
this case.
Using the reddening-corrected flux ratios \lineratzero\ of Table~\ref{tab02}, their
method yields logarithmic abundances of
   $12 + \log(\rm{O/H})  = 8.44\pm 0.02$ for oxygen and
   $12 + \log({\rm N/H}) = 7.55\pm 0.02$ for nitrogen.
Taken at face value, these abundances would indicate a relative logarithmic
gas metallicity of
   $[Z/\zsun] = -0.26\pm 0.02$
on the solar-abundance scale by \citet{aspl09}.
However, optical spectroscopy of a large sample of giant \hii\ regions in
{\em Spitzer}/SINGS galaxies by \citet{mous10} suggests that nebular
abundances derived through similar empirical calibrations need a logarithmic
correction of about $+$0.20 to be in accord with the \citeauthor{aspl09}
scale.
With such a correction, the logarithmic abundances for \hiisone\ become about
   $8.64\pm 0.05$ for oxygen and
   $7.75\pm 0.05$ for nitrogen, respectively,
and the relative gas metallicity $[Z/\zsun] \approx -0.06\pm 0.05$,
corresponding to a {\em linear} gas metallicity of
   $Z \approx 0.86\pm 0.10~\zsun$.
This new value agrees to within the combined errors with the value of
   $Z = 0.75\pm 0.08~\zsun$
found by \citet{ss98} through the traditional method \citep{mcga91,mcga94},
which included the \oii\ doublet.
Giving both quite independent measurements equal weight, we adopt the average
of
   $Z = 0.80\pm 0.07~\zsun$
for the metallicity of \hiisone.

In passing, we note that the reddening-corrected \halp-luminosity of
\hiisone\ within the measuring aperture is
   $L_0(\halp) = (1.13\pm 0.17)\times 10^{39}$ \ergsec,
from which we estimate a {\em total} true \halp-luminosity of
   $L_{0,{\rm tot}}(\halp) \approx (2$\,--\,$3)\times 10^{39}$ \ergsec.
This luminosity makes \hiisone\ comparable to 30 Doradus, the most luminous
giant \hii\ region in the Local Group, but still about one order of
magnitude less \halp-luminous than the most extreme extragalactic \hii\
regions \citep{sear71,kenn88}.

In summary, the giant \hii\ region surrounding Cluster S101 is highly
luminous, of slightly subsolar metallicity ($\sim${}0.80 \zsun), and
lies---based on its location and radial velocity (\S~\ref{sec34})---clearly
in the same stream of returning tidal material as the \oiiineb.
Assuming that its self-enrichment is relatively low, we adopt its
metallicity as likely being also that of the tidal material and nebula.

\begin{figure}
  \begin{center}
    \includegraphics[width=8.6cm]{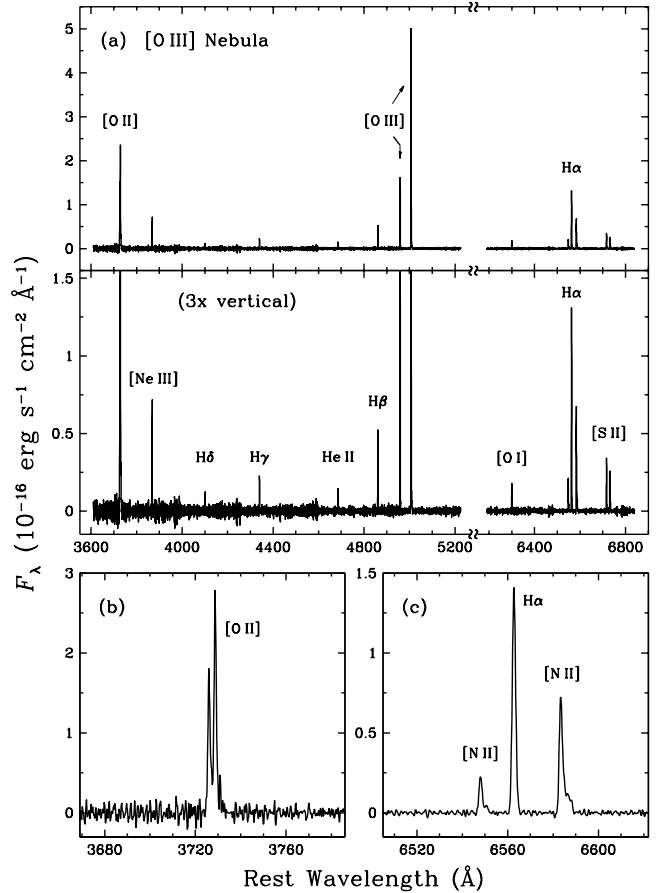}
    \caption{
MagE spectrum of the brightest part of the \oiiineb, plotted vs rest
wavelength.
This spectrum was obtained with the spectrograph slit positioned as shown in
Fig.~\ref{fig02}{\em a} (red bar) and covering a projected area of
$2.12\times 0.32$ kpc$^2$.
Sky and galaxy background have been subtracted.
(a) Slightly smoothed spectrum with emission lines identified and two
featureless stretches ($\lambda\approx 5200$\,--\,6200 and 6840\,--\,8120)
omitted; the smoothing was done with a Gaussian of $\sigma = 0.3$~\AA.
The flux scale of the lower panel is 3$\times$ expanded relative to that of
the upper panel.
Notice the strong \oiiboth\ and \oiiiboth\ lines as well as the presence
of \neiiieig\ and \heiisix\ lines.
Panels (b) and (c) display details of the \oii\ doublet and \halp\ + \nii\
lines, respectively, at full resolution (i.e., unsmoothed) and with a
10$\times$ expanded wavelength scale.
    \label{fig04}}
  \end{center}
\end{figure}

\subsection{Spectral Properties of the \oiii\ Nebula}
\label{sec33}

As described in \S~\ref{sec222}, a spectrum of the brightest part of the
\oiiineb\ was recorded with the MagE spectrograph through a $1\farcs0\times
10\arcsec$ slit, with the final extraction of the spectrum restricted to
a $6\farcs61$ long segment of the slit, corresponding to a projected area
of $2.12\times 0.32$~kpc$^2$ at the distance of \n7252.
The processed spectrum has a resolution of $\sim$4100 and covers the
wavelength range 3300\,--\,8250~\AA.
Figure~\ref{fig04} shows the two main segments of it that contain all
detected emission lines.
Details of two groups of lines are displayed in the bottom panels on a
10$\times$ expanded wavelength scale.
Notice the dominance of the \oiiboth\ and \oiiiboth\ lines.

The following subsections first describe the spectrum and our measurements
of emission-line fluxes from it, then derive estimates of the temperature,
electron density, and integrated \oiiisev\ and \halp\ line luminosities,
and finally discuss the position of the brightest part of the \oiiineb\
in several diagnostic line-ratio diagrams.

\subsubsection{Description of Spectrum}
\label{sec331}

The optical spectrum of the \oiiineb\ shown in Fig.~\ref{fig04} is distinctly
different from that of a normal giant \hii\ region (e.g., Fig.~\ref{fig03}),
with the main forbidden lines of oxygen, \oiiboth\ and \oiiiboth, clearly
dominant in flux.
It also shows the Balmer recombination lines \halp\,--\,\hdel\,\ and the
usual collisionally-excited lines \niiboth\ and \siiboth.
Perhaps most telling are the relatively high-excitation lines, besides
\oiiisev, of \neiiieig\ and \heiisix.
The latter requires full ionization of helium and, hence, an ionization
source rich in photons of energy $h\nu > 54.4$~eV.

Some diagnostic emission lines worth noting for their absence within the
noise limits of the MagE spectrum are \nevboth, \oiiithr, \niboth, \niifiv,
and \heisix.
The absence of the two \nev\ lines prevents us from commenting on the
possible presence of ionizing photons with energies of $h\nu > 97.1$~eV
(necessary to produce Ne$^{4+}$), but yields only weak constraints because
of the increased noise of the MagE spectrum at wavelengths $\la\,3600$~\AA\
(see \S~\ref{sec41}).

All emission lines measured in the \oiiineb\ are, with the exception noted
below, quite narrow.
This can be established from the MagE spectrum much better than from the
lower-resolution LDSS-3 spectra.
Gaussian fits to the line profiles of \oiiiboth, and \halp\ at six locations
along the MagE slit yield---after correction for the instrumental profile---a
 mean full width at half maximum of\ \ ${\rm FWHM} = 22.7 \pm 3.3$ \kms,
corresponding to a mean velocity dispersion in the ionized gas of
$\sigvelgas = 9.6\pm 1.4$ \kms.
This velocity dispersion is comparable to the isothermal sound speed in
ionized gas, $\sim$\,10 \kms\ at $T = 8000$ K \citep[e.g.,][]{spit78}.
Given the volume of the sampling regions, $0.45\times 0.32$ kpc$^2$ times the
depth of the \oiiineb\ along the line of sight, this low measured dispersion
suggests that {\em small-scale\,} velocity gradients within the nebula must
be relatively minor.

The one exception to the generally narrow line widths is the observed
asymmetric broadening of some forbidden lines near their base.
The strongest such broadening is observed for the \niiboth\ lines, as
illustrated in Fig.~\ref{fig04}{\em c}, whereas no similar broadening is seen
in the Balmer emission lines.
The cause of this asymmetric broadening is currently unknown.
Unfortunately, the signal-to-noise ratio of the MagE spectrum is
insufficient to permit reliable two-component fits for the various forbidden
lines and locations along the slit.

Fluxes of all emission lines were measured from the MagE spectrum shown
in Fig.~\ref{fig04} with the IRAF task {\em splot}.
Line profiles were fit with a single Gaussian or, when clearly broadened
near the base, with two Gaussians, and integrated.
Table~\ref{tab03} gives the resulting fluxes $F$ of the main, narrow
components.
Adding the flux of the broad components, when present and measurable, would
increase $F$ by $\la$\,5\% in all cases except for the \nii\ doublet, where
$F$ would increase by about 20\% to 25\% for both lines.
Even the latter sizable flux increase would, however, not significantly
affect any of the results derived in the present paper.
The upper flux limits given in Table~\ref{tab03} for four interesting, but
undetected emission lines, \nevfou, \oiiithr, \niifiv, and \heisix,
correspond to $2\,\sigma$ of the local noise. 

Table~\ref{tab03} also gives flux ratios relative to \hbet\ corrected for Milky-Way
foreground reddening, \lineratzeromw, and for Milky-Way plus
internal reddening, \lineratzero.
The foreground-reddening corrections were computed from the adopted
$\av = 0.100$ \citep{schl98} using the reddening curve by \citet{card89},
as approximated by \citeauthor{odon94}'s (1994) formulae.
After correcting the observed Balmer decrement for foreground reddening, the
internal reddening was estimated by comparing the decrement with that of a
Case B spectrum for $T = 9000$~K and low density \citep{of06}, but with the
ratio \halp/\hbet\ set to 3.1.
This slightly increased value has been shown to be appropriate for nebulae
photoionized by a power-law source \citep{halp83,gask84}, which seems to be
the case here (see \S~\ref{sec333}).
The resulting weighted mean internal extinction, $\avint\approx 0.16\pm
0.04$, is only a rough estimate since it rests on the simplistic assumption
that the dust distribution is screen-like.
Hence, we adopt a more conservative value of $\avint = 0.15 \pm 0.15$ that
indicates the entire likely range of internal visual extinction.
The flux ratios corrected for both Milky-Way and internal reddening,
\lineratzero, are given in the last column of Table~\ref{tab03}.

\subsubsection{Electron Density, Temperature, and Line Luminosities}
\label{sec332}

From the line fluxes listed in Table~\ref{tab03}, we can estimate the mean electron
density \nel\ and temperature $T$ in the part of the \oiiineb\ covered by
the MagE slit.

The mean line ratio of the \oii\ doublet, $F(\lambda 3729)/F(\lambda 3726) =
1.55\pm 0.06$, yields $\nel \la 10$ cm$^{-3}$ \citep[e.g.,][]{of06}, while
(noisy) individual values of the ratio along the slit suggest a range of\,
$1 \la \nel \la 300$ cm$^{-3}$, with values above 100 cm$^{-3}$ rare.
The mean line ratio of the \sii\ doublet, $F(\lambda 6716)/F(\lambda 6731) =
1.32\pm 0.08$, yields $\nel = 100^{+100}_{-70}$ cm$^{-3}$.
Thus, within the area covered by the MagE slit, the electron density seems to
be generally low, $\nel \la (2$\,--\,$3)\times 10^2$ cm$^{-3}$.

We can gain some additional insight into the density distribution within the
nebula from the two single-slit spectra obtained with LDSS-3 (\S~\ref{sec221}).
The blue spectrum does not cover the \oii\ doublet, but the red spectrum
does include the diagnostic \sii\ doublet and shows that
$F(\lambda 6716)/F(\lambda 6731) > 1.0$ along the entire section of the
\oiiineb\ covered by the slit (see Fig.~\ref{fig02}).
This limits the density to $\nel < 500$ cm$^{-3}$ along that $\sim$10~kpc
long section with---again---most \nel\ values significantly lower than this
upper limit.
Thus, it seems reasonable to assume that the entire visible \oiiineb\ has
a large-scale ($\ga$\,0.5 kpc) mean \nel\ significantly below this limit.
Most diagnostic line ratios can, therefore, be interpreted within their
low-density approximations (\S~\ref{sec333}).

Upper limits for the {\em mean temperature} of the \oiiineb\ within the
MagE aperture can be estimated from \oiii\ and \nii\ line ratios.
Using the fully reddening-corrected fluxes of Table~\ref{tab03}, the \oiii\ line
ratio\,\ $\{F_0(\lambda 4959)+F_0(\lambda 5007)\}/F_0(\lambda 4363) > 323$\ \
yields\,\ $T < 8860$~K for $\nel < 500$ cm$^{-2}$
\citep[Equation (5.4)]{of06}.
Similarly, the \nii\ ratio\,\
$\{F_0(\lambda 6548)+F_0(\lambda 6583)\}/F_0(\lambda 5755) > 86.2$\,\
yields\,\ $T < 10,550$~K (ibid., Equation (5.5)).
In what follows we adopt the stricter, rounded-up limit of\,\ $T < 8900$~K
for the nebular temperature.

Integrated line luminosities for the entire \oiiineb\ are of interest since
they may help establish the source and history of excitation.
Here we derive the integrated \oiiisev\ and \halp\ luminosities as follows.

\begin{figure*}
  \begin{center}
    \includegraphics[width=17cm]{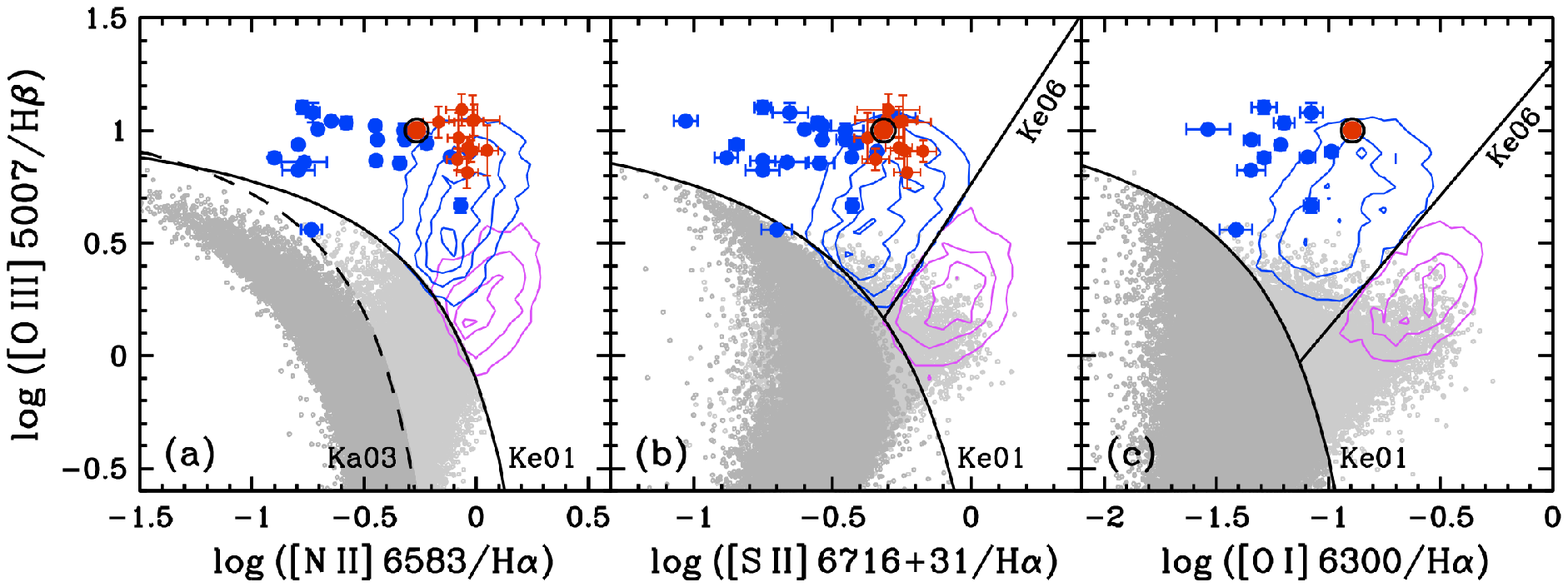}
    \caption{
Position of \n7252's \oiiineb\ (red dots) in three standard line-ratio
diagrams that plot the ratio \oiiisev/\hbet\ versus the ratios
(a) \niithr/\halp, (b) \siiboth/\halp, and (c) \oizer/\halp.
The dashed curve in panel (a) marks the empirical upper boundary for purely
star-forming galaxies \citep{kauf03}, while the solid curves in each panel
mark the theoretical upper boundary for extreme starbursts determined by
\citet{kewl01}.
Straight lines in panels (b) and (c) mark the empirical boundaries between
LINERs (below) and Seyfert galaxies (above) determined by \citet{kewl06}.
For reference, the line ratios of a subsample of emission-line galaxies from
the SDSS Data Release 4, selected for high signal-to-noise line measurements,
are shown in gray for star-forming (dark) and composite-spectrum (light)
galaxies, and as contours for LINERs (red) and Seyfert galaxies (AGNs, blue).
For comparison with \n7252, the line ratios of various EELRs around five QSOs
measured by \citet{fust09} are shown as blue data points with error bars.
Our measurements of the \oiiineb\ made with MagE are shown as a large red dot
with black rim in each panel (error bars are smaller than the dot), while
those made with LDSS-3 along its long slit are shown as small red dots with
error bars.
The \oiiineb\ clearly lies in the domain of Seyfert galaxies, indicating a
power-law source of excitation.
    \label{fig05}}
  \end{center}
\end{figure*}

The \oiiisev\ flux measured within a $6\farcs61\times 1\farcs0$ subsection
of the MagE slit is $(7.68\pm 0.19)\times 10^{-16}$ \ergcms\ (Table~\ref{tab03}).
We used it to calibrate the counts in our continuum-subtracted, but otherwise
uncalibrated \oiiisev\ image (\S~\ref{sec21} and Fig.~\ref{fig01}).
Note that the \oiiisev\ line is the only emission line of the nebula
within the narrow passband used for the exposure.
Integrating counts over an area of $31\arcsec\times 19\arcsec$
($\sim$\,$10\times 6.1$~kpc$^2$) oriented at P.A.\,= $103\fdg3$, which covers
all measurable \oiii\ line emission of the nebula, and converting the sum to
flux units yields a total observed flux of
$F(\lambda 5007) = (2.02\pm 0.07)\times 10^{-14}$ \ergcms.
The total {\em reddening-corrected}\,\ flux is
$F_0(\lambda 5007) = (2.62\pm 0.48)\times 10^{-14}$ \ergcms, and the
corresponding true integrated line luminosity of the \oiiineb\ at the
distance of \n7252 is
$L_0(\lambda 5007) = (1.37\pm 0.25)\times 10^{40}$ \ergsec.
Note that the relatively large errors for the last two, reddening-corrected
quantities reflect mainly the uncertainty in our adopted value for the
internal reddening, $\avint = 0.15\pm 0.15$ (\S~\ref{sec331}).

The determination of the integrated \halp\ flux and luminosity is slightly
more complicated since the \niithr\ line enters the narrow passband used
for the ``\halp'' imaging at nearly full strength, while the \niieig\ line
lies on the blue flank of the passband and enters at about 44\% of peak
transmission.
The \halp\ line itself is located at peak transmission.
Hence, to calibrate the counts of the ``\halp'' image in flux units we
have to calculate the weighted sum of the three line fluxes given in
Table~\ref{tab03}, $F(\halp) + 0.96\,F(\niithr) + 0.44\,F(\niieig)$, and divide it by
the integrated number of image counts within the $6\farcs61\times 1\farcs0$
subsection of the MagE slit.
On the assumption that the $\nii/\halp$ line ratios within the slit are
representative of those across the entire \oiiineb, we can then estimate
the integrated \halp\ flux of the entire nebula over the same $31\arcsec
\times 19\arcsec$ area used for the \oiiisev\ line above; we find a total
observed flux of $F(\halp) = (6.60\pm 0.62)\times 10^{-15}$ \ergcms.
The total {\em reddening-corrected} flux is
$F_0(\halp) = (7.93\pm 1.66)\times 10^{-15}$ \ergcms, and the corresponding
true integrated \halp\ luminosity of the nebula is
$L_0(\halp) = (4.16\pm 0.87)\times 10^{39}$ \ergsec.
The $\sim$\,21\% errors of the last two quantities reflect the combined
uncertainties of the background subtraction, flux calibration, and internal
reddening correction.

As a check on our integrated luminosities, the ratio
$L_0(\oiiisev) / L_0(\halp) = 3.29 \pm 0.30$  agrees well with the flux ratio
measured within the MagE subaperture (Table~\ref{tab03}),
$F_0(\lambda 5007)/ F_0(\halp) = 3.25 \pm 0.17$.
The fluxes measured with MagE were, of course, used to calibrate the
continuum-subtracted narrow-band images, yet it is reassuring to know
that the continuum subtractions and count integrations performed on these
images did not significantly change the ratio.
The good agreement also supports the assumption that the relevant line ratios
do not vary much across the entire \oiiineb\ (see also \S~\ref{sec333} below).

\subsubsection{Line-Ratio Diagnostics}
\label{sec333}

Ratios of emission-line fluxes from ionized-gas nebulae provide important
diagnostic information concerning the possible source, or sources, of
excitation \citep[e.g.,][]{bald81,vo87,kewl01,kewl06}.

Figure~\ref{fig05} shows reddening-corrected logarithmic line ratios
for the \oiiineb\ (red dots) plotted in three standard diagnostic
line-ratio diagrams.
Also plotted, for comparison, are line ratios for a number of extended
emission-line regions (EELRs) around quasars observed by
\citet[blue dots]{fust09} and for about 102,000 emission-line galaxies
from the Sloan Digital Sky Survey (SDSS, gray dots plus red and blue
isopleths).
Curves and straight lines mark well-known boundaries for starforming,
composite-spectrum, LINER, and Seyfert galaxies, as detailed in the
caption.
As the high-quality line ratios measured with MagE (black-circled red dot
in each panel) show, the \oiiineb\ clearly lies in Seyfert territory
(blue isopleths), well above the boundary for extreme starbursts
\citep{kewl01} and away from the dividing line between Seyfert and LINER
galaxies \citep{kewl06}.
Emission-line nebulae in that part of the diagrams, especially at
$\log (\oiiisev/\hbet) \ga 0.9$, are most often photoionized by a source
with a power-law spectrum.

According to photoionization model grids by \citet{grov04} shown in
\citet[Fig.~11b]{fust09} the \oiiineb's position in the
metallicity-sensitive \oiii/\hbet\ vs \nii/\halp\ diagram suggests
gas of slightly sub-solar metallicity ($\sim$\,$2/3$ \zsun)
photoionized by a power-law-spectrum source with an exponent of
about $-1.5$.
Supporting this tentative interpretation is the fact that our measured
abundance of $Z = 0.80 \pm 0.07$ \zsun\ in the nearby giant HII region
\hiisone\ (\S~\ref{sec32}) yields a similarly sub-solar abundance
for the gas in that part of \n7252's tidal tail.
Furthermore, the line ratios of the \oiiineb\ lie close to those of the
EELRs around quasars measured by \citet{fust09}, all of which appear to be
of subsolar metallicity and are clearly photoionized by the central engine.

\begin{figure}
  \centering
  \includegraphics[width=8.5cm]{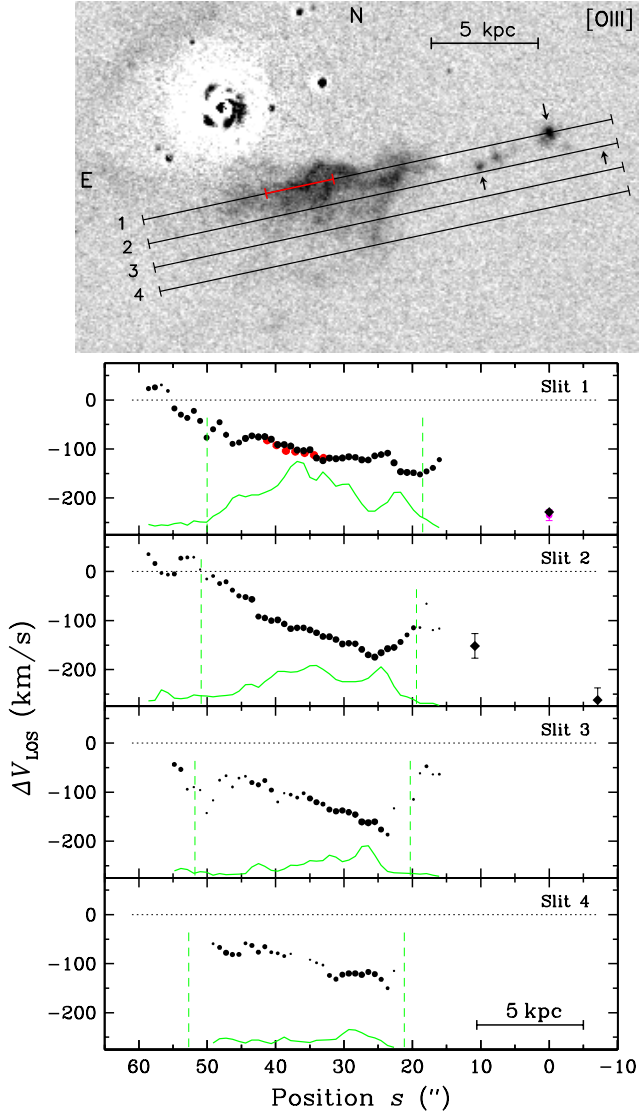}
  \caption{
Slit positions and corresponding ionized-gas velocities measured across
\n7252's \oiiineb.
({\em Top}) Continuum-subtracted \oiiisev\ image showing the positions of
slits used with the LDSS-3 and MagE spectrographs.
Black lines mark the four LDSS-3 slits (each 70\arcsec\ long) used to measure
the velocities shown in the diagram underneath.
The short red line marks the position of the MagE slit also used.
Arrows point to three giant \hii\ regions with measured velocities plotted
below; the downward arrow points to \hiisone, whose central star cluster
served as a {\em spatial} reference point for the measurements.
({\em Bottom}) Line-of-sight velocities of the ionized gas relative to the
systemic velocity of the galaxy ($+$4749 \kms).
Black data points show \oiiisev\ velocities measured with LDSS-3; larger
points indicate more reliable measurements.
In the Slit 1 panel, red points display measurements made with MagE, while
the magenta point with error bar at $s = 0\arcsec$ (partially covered by a
black point) marks the velocity of \hiisone\ measured by \citet{s82}.
Thin green profiles show the \oiiisev\ flux on a uniform, but arbitrary
linear scale.
For details, see text.
  \label{fig06}}
\end{figure}

To check on our MagE results, we also measured line ratios from the
blue and red LDSS-3 spectra along the $\sim$10 kpc cross section of the
nebula covered by the main long slit (Fig.\ 2 and \S~\ref{sec221}).
Figures~\ref{fig05}{\em a} and \ref{fig05}{\em b} display these ratios as
red dots with error bars.
(Note that no $\oi/\halp$ ratios are shown since the \oizer\ line lies too
close to the ends of the blue and red spectra to be measured reliably.)
The ratios measured with LDSS-3 support those measured with MagE, and
hence the above conclusions concerning the metallicity and likely
ionization mechanism.
The one minor exception is that the $\nii/\halp$ ratios measured with LDSS-3
exceed those measured with MagE by $\sim$70\% on average.
About 10\% of this excess can be attributed to the weaker \halp\ flux
measured with LDSS-3, where the relatively low resolution makes the
subtraction of the underlying continuum with \halp\ absorption
considerably more difficult.
Also, the broad components of the \nii\ doublet mentioned above
(\S~\ref{sec331}) are unresolved in the LDSS-3 spectra, adding
20\%\,--\,25\% to the measured flux (see Fig.~\ref{fig04}{\em c}).
Hence, about half of the $\nii/\halp$ excess measured with LDSS-3
relative to the higher-resolution MagE spectrum is explained, but
the other half remains unexplained.
Even with this uncertainty, however, the LDSS-3 line ratios confirm the
main MagE result that \n7252's \oiiineb\ is likely photoionized and
possibly related to EELRs seen around quasars.

Finally, we searched for any possible gradients in the three line ratios
$\oiii/\hbet$, $\nii/\halp$, and $\sii/\halp$ that might indicate
varying excitation levels along the cross section covered by the LDSS-3
long slit and found none.
However, remember that in the various filaments of the nebula extending
from its bright ridge southward the \oiii\ and \halp\ images
(Figs.~\ref{fig01}{\em e} and \ref{fig01}{\em f}) strongly suggest the
presence of excitation ($\oiiisev/\halp$) variations, as described in
\S~\ref{sec31}.
Hence, the question of such variations across the {\em entire} nebula will
need to be reexamined through detailed observations with an
integral-field-unit (IFU) spectrograph.

\subsection{Kinematics of the \oiii\ Nebula}
\label{sec34}

The kinematics of the \oiiineb\ holds clues about both the nebula's relation
to the gas of \n7252 surrounding it and its own internal motions.
We used the blue 2D spectrum obtained with LDSS-3 through a mask with
four parallel slits (\S~\ref{sec221}) to roughly map this kinematics and the
velocities of three giant \hii\ regions in the W loop.
The results are shown in Figure~\ref{fig06}, where black data points in the
bottom four panels represent line-of-sight velocities \dvlos\ relative to the
systemic velocity of the galaxy, $\czsyshel = +4749$ \kms\ \citep{s82}.

These velocities were measured across the \oiiineb\ with the IRAF task 
{\em splot} and an ``aperture'' of $1\farcs7$ (9 pixels) in steps of
$0\farcs95$ (5 pixels) along each of the four slits.
Although emission lines other than \oiiisev\ were also measured when
feasible, only the $\lambda$5007 line was strong enough to be measured
consistently along most of each slit.
Hence, the plotted data points for the nebula and ionized gas surrounding it 
represent measurements from only this line, their sizes being proportional
to the estimated quality of the measurement.
For the largest points, the estimated $\pm\,1\sigma$ error is roughly
of the size of the plotted point ($\pm5$ \kms), while the smallest, least
reliable data points may have errors of up to about $\pm25$ \kms.
Note that along Slit 1 the velocities measured with LDSS-3 (black data
points), where they overlap with those measured with MagE (red data points),
agree excellently.
Solid green lines in each of the four bottom panels of Figure~\ref{fig06}
represent the \oiiisev\ line flux on an arbitrary, but uniform scale, and
pairs of vertical green dashed lines mark the two right-ascension boundaries
of the nebula, $\Delta\alpha = -31\farcs5$ and $0\farcs0$\,\ relative to the
nucleus (\S~\ref{sec21}).
As the figure shows, the general trend of \dvlos\ is to {\em decrease} (i.e.,
become more negative) from E to W across the nebula, with most relative
velocities ranging between about $-$50~\kms\ and $-$200~\kms.

\begin{figure}
  \begin{center}
    \includegraphics[width=8.4cm]{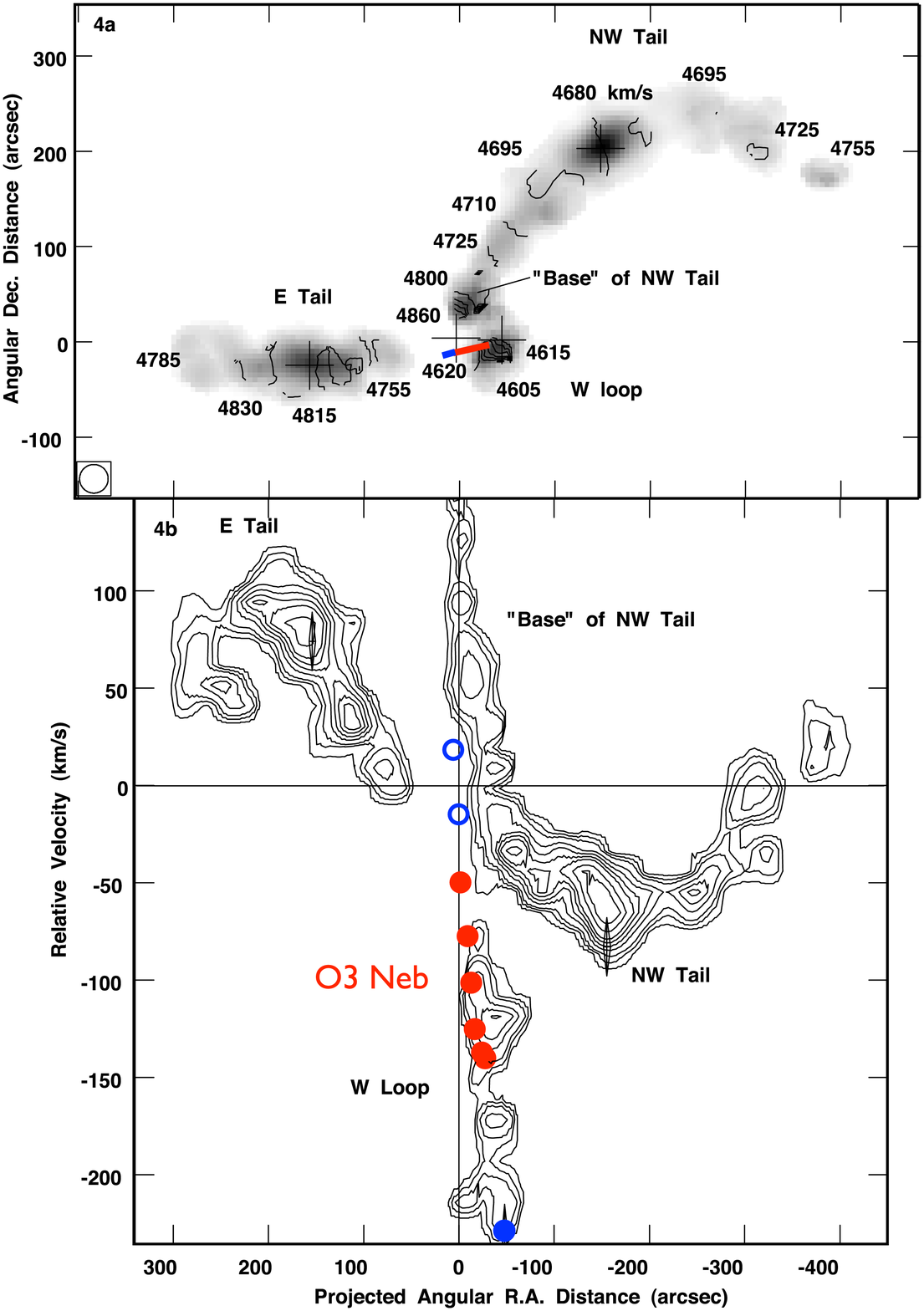}
    \caption{
Line-of-sight velocities of ionized gas in and near the \oiii\ nebula 
compared with those of neutral hydrogen associated with the tails and W loop
of \n7252, as mapped by \citet[Fig. 4]{hibb94} with the Very Large Array.
(Top) The projected \hi\ distribution as seen on the sky (gray-shaded areas
with superposed velocity contours) and locations where ionized-gas velocities
were measured with LDSS-3 (blue-and-red bar).
The cross above the bar marks the location of the nucleus, while that to its
right marks the location of \hiisone.
(Bottom) Position--velocity diagram; LOS velocities relative to the galaxy's
systemic velocity are plotted vs $\Delta$RA (in arcsec).
Contours mark \hi\ profiles obtained by integration along the declination
axis; components stemming from the E tail, W loop, and NW tail are labeled.
Colored data points show mean ionized-gas velocities measured in the
\oiiineb\ (red), in gas to its east (blue open), and in \hiisone\ 
(blue filled).
As detailed in \S~\ref{sec34}, the combined velocity data strongly suggest
that the \oiiineb, \hii\ regions to its west, and \hi\ are all part of
the gas stream forming the W loop and falling back into the remnant.
(Underlying figure courtesy of John E.\ Hibbard.)
    \label{fig07}}
  \end{center}
\end{figure}

As Figure~\ref{fig06} also shows, this trend seems to continue beyond the
\oiiineb\ into the ionized gas of the W loop, where Slit 1 crossed the giant
\hii\ region \hiisone\ by design and Slit 2 crossed two \hii\ regions (marked
in the top panel by upward-pointing arrows) at least in their outskirts by
chance.
The velocities of these three regions were measured from multiple emission
lines to improve their accuracy.
Note that our new value of \dvlos\ for \hiisone, $-228.5 \pm 2.1$ \kms,
agrees well with that measured by \citet{s82}, $-235 \pm 11$ \kms\ after
correction for redshift stretching ignored by him (magenta point with error
bars); we adopt a weighted average of $\dvlos = -230 \pm 5$ \kms\ for
this giant \hii\ region.
In contrast, the \dvlos\ of the two \hii\ regions crossed by Slit 2,
$-152\pm 25$ \kms\ and $-262\pm 25$ \kms, have relatively large errors because
these regions were not centered on the slit, leading to non-uniform slit
illuminations unknown in detail.
Taken all together, these optical velocity measurements strongly support
the notion that {\em the \oiiineb\ lies in a stream of at least partially
ionized gas that includes the southern branch of the W loop.}

It is interesting to also compare these ionized-gas velocities with the
neutral-gas \hi\ velocities measured by \citet{hibb94}.
Figure~\ref{fig07} shows, at the top, the \hi\ distribution in \n7252 with
the locations of our ionized-gas velocity measurements indicated by a
blue-and-red bar.
To approximate the lower spatial resolution of the \hi\ observations, we
averaged the measured ionized-gas velocities from slits 1 and 2
(Fig.~\ref{fig06}) in 5\arcsec\ long segments along the slits, yielding six
mean velocities within the \oiiineb\ plus two to its east.
The bottom panel of Fig.~\ref{fig07} shows the six velocities of the
\oiiineb\ as solid red dots superposed on a position--velocity diagram
of the \hi\ velocities displayed in contour form (for details, see
\citealt[Fig.\ 4]{hibb94}).
The two blue open circles mark ionized-gas velocities to the east of
the \oiiineb, while the filled blue dot near the bottom of the diagram
marks the velocity of \hiisone.
From these various data points, the \oiiineb\ clearly appears to be associated
with the \hi\ in the W loop.
In contrast, the ionized gas to its east (blue circles) appears more likely
to be associated---if at all---with the NW tail than with the E tail.
Further, the giant \hii\ region \hiisone\ clearly appears embedded in the
\hi\ of the southern branch of the W loop.

In summary, the combined velocity data strongly suggest that the \oiiineb,
the various \hii\ regions to its west, and the neutral \hi\ are all
part of the gas stream forming the southern branch of the W loop.
This gas stream is clearly falling back into the remnant.
Whether it itself is a continuation of the gas stream in the E tail, as the
\hi\ profiles of Fig.~\ref{fig07} and dynamical models of \n7252
\citep{himi95,chba10} suggest, is at present less certain.
We comment on this issue further in \S\S~\ref{sec43} and
\ref{sec46} below.

\section{DISCUSSION}
\label{sec4}

The present discussion focuses first on the excitation mechanism for the
\oiiineb, which most likely is photoionizaton by a power-law source in the
nucleus of \n7252.
It then describes the lack, so far, of any detection of AGN activity in
this remnant, with tight upper limits.
We then put forth the hypothesis that the nebula is an ``ionization echo''
similar to others recently discovered in some weakly active galaxies,
except that is it over an order of magnitude less luminous.
After a brief discussion of the implications for the history of activity
in \n7252, we point out some remaining questions and suggest further work.

\subsection{Excitation Mechanism for \oiii\ Nebula}
\label{sec41}

The position of the \oiiineb\ in the three line-ratio diagrams shown in
Figure~\ref{fig05} and the presence of \heiisix\ line emission in its
spectrum (Fig.~\ref{fig04}) strongly point toward the nebula beeing
{\em photoionized}\, by a source with a relatively hard spectrum rich in
$h\nu > 54.4$~eV photons (\S~\ref{sec331}).

Yet, given the nebula's location in the W loop's gas stream that is falling
back into the remnant, possibly from the E tidal tail (\S~\ref{sec34}), one
has to seriously consider the possibility that some {\em shocks} might
occur in that stream, whether it be through collision of the stream with
other gas in the already rather relaxed halo of \n7252, with an outflow
from the central region, or perhaps even with a so far undetected jet
emanating from the nucleus.

The three main reasons why we think shock excitation of the \oiiineb\ is
unlikely are as follows.
First, shock excitation typically yields high electron temperatures,
$T \ga 15,000$, yet our measured upper temperature limit of $T \la 8900$
for the nebula indicates much cooler gas.
The few ``shock\,+\,precursor'' models for near-solar abundances in the
MAPPINGS III library of shock models \citep{alle08} that do reach
temperatures below 10,000 K have shock velocities well below 300 \kms,
quite insufficient to produce the observed high excitation.
Second, only models with shock velocities of $\vshock \ga 450$ \kms\ get
close to being able to reproduce the observed line ratio of
$\oiiisev/\hbet = 10.03\pm 0.25$
(Table~\ref{tab03}), yet the measured emission-line widths indicate a low velocity
dispersion in the gas of $\sigvelgas = 9.6\pm 1.4$ \kms, a value close to
the isothermal sound speed in ionized gas of the measured temperature
(\S~\ref{sec331}); it is difficult to conceive how the above high shock
velocities could be present without creating significant turbulence and
line broadening.
And third, the observed line ratio of $\heiisix/\hbet = 0.32 \pm 0.02$
(Table~\ref{tab03}) is difficult to reproduce with shock velocities of less than
about 350 \kms\ and, hence again, incompatible with the low measured
temperature.

An alternative way to summarize these difficulties is that in diagnostic
diagrams with shock-sensitive line ratios the MAPPINGS III models can easily
be made to cover the domain of LINERs, but not the domain of Seyfert galaxies
(e.g., Figs.\ 33 and 34 in \citealt{alle08}), in which the \oiiineb\
clearly lies.
In short, the line strengths measured in the nebula seem to strongly favor
photoionization by a relatively hard source of radiation over some form of
internal shock excitation.

Accepting photoionization by a presumably central power-law source with
a relatively hard spectrum as the likely excitation mechanism, we can
estimate the mean ionization parameter $U$ in the brightest part of the
\oiiineb\ (covered by the MagE slit) from the reddening-corrected ratio
$\log (\oii \lambda 3727/\oiiisev) = -0.086 \pm 0.018$,
using the photoionization model by \citet{komo97} and the approximation
formulae by \citet[][esp.\ Appendix C.4]{benn05}.
The logarithmic value of this parameter, which represents the dimensionless
ratio of the local ionizing photon density to the electron density
\citep[e.g.,][]{of06}, is
$\log U \approx -2.9$ for $\nel \approx 10^2$ cm$^{-3}$ and
$\log U \approx -2.7$ for $\nel \approx 1$ cm$^{-3}$.
These values lie well within the range of $-3.5\ga \log U\ga -1.5$ typically
measured in EELRs associated with quasars and Seyfert galaxies
\citep[e.g.,][]{fust09,keel12a}.

Note that our non-detection, in the \oiiineb, of the \nevfou\ line often
seen in EELRs does not weaken any of the above conclusions.
The measured upper limit for this line's flux ratio to \hbet, 
$F_0(\nevfou)/F_0(\hbet) < 0.29$ (Table 3), places the \oiiineb\ among
relatively week \nev-line emitters in the sample of 19 EELRs (``AGN-ionized
clouds'') studied by \citet{keel12a}.
However, this limit lies still well within the range of
$0.044 \leq \nevfou/\hbet \leq 2.43$ observed for the sample, with seven
EELRs falling below the limit for the \oiiineb.
Similarly, the upper limit for the \nev\ line's flux ratio relative to
\neiiieig, $F_0(\nevfou)/F_0(\neiiieig) < 0.24$, also places the \oiiineb\
potentially within the range of $0.06 \leq \nevfou/\neiiieig \leq 2.64$
measured for the sample of EELRs \citep[Table 4 of][]{keel12a}, in this case
above the bottom quintile (four EELRs out of 19).
Clearly, detecting and measuring the \nevfou\ line in the \oiiineb\ will
not be easy because of its low surface brightness, but it will be important
for checking our above conclusion that the nebula is most likely
photoionized by a central power-law source in \n7252.

\vskip 0.7truecm
\subsection{Properties of the Nucleus}
\label{sec42}

Given the above evidence that the \oiiineb\ is, or has been, excited by a
power-law source, it seems perplexing that no clear evidence of any
nuclear activity in \n7252 has been found to date.

{\em Optically}, the spectrum of the nucleus shows strong Balmer absorption
lines typical of a post-starburst stellar population, with only relatively
weak emission lines superposed \citep{s82,schw90,liu95}.
Table~\ref{tab04} gives the \oiisev, \hbet, \oiiisev, and \halp\ flux ratios and
apparent luminosities measured with the MagE spectrograph through a
$0\farcs9\times 0\farcs7$ ($289\times 225$ pc$^2$) aperture centered on the
nucleus and, for comparison, the corresponding flux ratios and luminosities
for both the central ionized-gas disk out to $r = 4\arcsec$ (1.28 kpc)
and the entire \oiiineb.
The data for the nucleus and central gas disk are taken from a detailed
study of the central region of \n7252 to be published separately (Schweizer
et al., in preparation).
Note that the apparent \halp\ emission-line luminosity of the nucleus,
computed from the
measured line flux on the assumption of negligible internal extinction,
is a measly $\lappnuc(\halp) = (1.63\pm 0.40)\times 10^{39}$ \ergsec,
whereas that of the surrounding ionized-gas disk within
$0\farcs5 < r \leq 4\arcsec$ (0.16\,--\,1.28 kpc) is
$\lappdisk(\halp) = (2.54\pm 0.25)\times 10^{41}$ \ergsec, or a good two
orders of magnitude higher.
Similarly, the apparent \oiiisev\ luminosity of the nucleus is only
$\lappnuc(\lambda 5007) = (2.3\pm 0.9)\times 10^{38}$ \ergsec, while that
of the ionized-gas disk is about 40$\times$ higher.
Note also that the ratio of this nuclear \oiiisev\ luminosity to that of
the entire \oiii\ nebula is small,
$\lappnuc(\lambda 5007)/\lappneb(\lambda 5007) = 0.017$.
In short, the {\em nuclear} optical-line emission appears weak and is nearly
negligible compared to the line emission from the surrounding, strongly
star-forming ($\sim\,$6 \msunyr) central gas disk; it definitely does not
show any signs of nuclear activity.

Could the nucleus be highly reddened and its emission-line fluxes strongly
extincted?
The surface-brightness and color profiles measured with the {\em Hubble
Space Telescope (HST)}\, \citep{lain03,ross07} do not suggest so.
On the contrary, the \vk\ profile, which stays approximately flat at
$\vk \approx 3.25$ from $r = 0\farcs4$ to 5$\arcsec$, turns slightly
{\em bluer} within $r < 0\farcs4$, reaching $\vk = 3.13$ in the central
$r < 0\farcs1$.
And the central surface-brightness profile appears as a pure power law
with essentially the same index in both $V$ \citep{mill97} and $K$
\citep{ross07}.
Hence, there is no sign of significant nuclear reddening or extinction.
This is consistent with the measured flux ratio
$\fzeromw(\halp)/\fzeromw(\hbet) = 3.1\pm 1.2$ corrected only for Milky-Way
foreground reddening (Table~\ref{tab04}), though the large uncertainty makes
this a relatively weak check.

In the {\em mid-infrared}, a significantly stronger check on the possible
presence of any highly reddened AGN is provided by two magnitudes measured
with the {\em Wide-field Infrared Survey Explorer (WISE)} \citep{wrig10},
$W1$ at $\lambda\approx 3.4\,\mu$m and $W2$ at $\lambda\approx 4.6\mu$m.
The simple color criterion $W1\!-\!W2 \geq 0.8$, in connection with the
magnitude cutoff $W2 \leq 15.0$, reveals AGNs both unobscured and obscured
over the redshift range $0 < z \la 2.5$ with great efficiency
\citep{ster12}.
Yet the central magnitudes measured for \n7252 with a point-spread function
of $\sim\,$6\arcsec\ FWHM, $W1 = 9.793$ and $W2 = 9.596$ \citep{cutr12},
yield a color of $W1\!-\!W2 = 0.20\pm 0.03$, placing the nucleus far short of
the realm of local Seyfert galaxies ($W1\!-\!W2\approx 0.8$ to 1.7 mag).
Hence, there is no evidence in the mid-infrared either for any strongly
reddened active nucleus in \n7252, in agreement with a more detailed
{\em Spitzer}/IRS spectroscopic study by \citet{bran06}.

There is also no evidence from the {\em radio continuum} for any
significant nuclear activity.
A central radio-continuum source has been detected with a flux density
of $19\pm 0.5$ mJy at 1.4 GHz and ``is possibly resolved'' when observed
at a resolution of $15\arcsec\times 15\arcsec$ (FWHM) with the VLA's hybrid
CnB array \citep{hibb94}.
At the distance of \n7252, this flux density corresponds to a spectral
luminosity of\,\ $\lradio = (1.00\pm 0.03)\times 10^{29}$ \ergsechz, which
in turn indicates a star formation rate (SFR) of $6.3\pm 0.2$ \msunyr\ (via
the conversion relation by \citealt{murp11}).
This extinction-free estimate of the SFR agrees well with the SFR of
$5.6\pm 1.1$ \msunyr\ determined from the integrated \halp\ luminosity
of the ionized-gas disk (corrected for an internal extinction of
$\av = 1.75\pm 0.25$ mag; Schweizer et al., in prep.).
Hence, there is no need for any {\em nuclear} contribution to the observed
\lradio, and we estimate an upper limit for any such contribution to be
less than 10\% of the detected luminosity, or\,\
$\lradionuc < 1.0\times 10^{28}$ \ergsechz.
This radio power would still permit the nucleus of \n7252 to lie in the
mid-range of radio powers of local Seyfert galaxies
($\sim$\,$4\times 10^{27}$ \ergsechz\ for detected Seyferts; \citealt{lho01}),
but would make \n7252 at least an order of magnitude less radio loud than
the well-known nearby Seyferts \n1068 (Sey2, $3.1\times 10^{29}$ \ergsechz)
and \n4151 (Sey1.5, $8.5\times 10^{28}$ \ergsechz).
Obviously, better than $\sim$\,1$\arcsec$ resolution radio-continuum images
will be needed to address the question whether \n7252's nucleus shows any
detectable radio emission.

Finally, if there is any hard {\em X-ray emission} from the nucleus
betraying its activity, it has to be quite weak.
The suggestion by \citet{awak02} that the 2.0\,--\,10 keV X-ray emission
detected with ASCA's Solid-State Imaging Spectrometer (SIS) may indicate
nuclear activity in \n7252 is questionable \citep{nola04}, given that
the 6$\arcmin$ diameter region used for the spectrum extraction contains
the entire body of the remnant, including its numerous known extra-nuclear
X-ray sources.
As \citet{nola04} show from {\em Chandra} observations, the relatively
X-ray luminous ``nuclear'' source is clearly extended, with a break in
its radial counts profile occurring at $r\approx 3\farcs0$ (where the
SFR in the central gas disk drops off sharply).
An analysis of its {\em XMM--Newton} spectrum in terms of gaseous and
absorbed-power-law components reveals the presence of a possible power-law
component peaking at 2.0 keV and traceable over the energy range
1.4\,--\,4.0 keV.
The absorption-corrected luminosity of this component is
$\lxzero = 2.10\times 10^{40}$ \ergsec, with a 90\% confidence interval
of (1.27\,--\,$2.25)\times 10^{40}$ \ergsec\ (values adjusted to our
distance of 66.2 Mpc adopted for \n7252).

Yet, the unabsorbed 2\,--\,10 keV luminosity expected from the intense
star formation ($\sim$\,$6.3\pm 0.2$ \msunyr) in the central gas disk
is $\lxttzero = (3.7\pm 0.1)\times 10^{40}$ \ergsec, based on the
relation
$$ {\rm SFR}\ (\msunyr) = 1.7\times 10^{-40}\ \lxttzero\ (\ergsec) $$
determined by \citet{rana03} and adjusted to the \citet{krou01} initial
mass function (IMF) by \citet{kenn12}.
Hence the entire measured X-ray luminosity of the central hard component
is more than fully supplied by the star-forming central disk, and there
is no need to invoke any extra contribution from an AGN.
We estimate that any such hypothetical nuclear X-ray component, if it
exists, has currently a luminosity of\,\ $\lxttzero < 5\times 10^{39}$
\ergsec.

In summary, the current level of nuclear activity in \n7252 appears to
be low from X-ray through optical to radio wavelengths and remains---if
it exists---undetected so far.

\vskip 0.8truecm

\subsection{The \oiii\ Nebula as a Possible ``Ionization Echo''}
\label{sec43}

The observed high excitation level of the \oiiineb\ in \n7252 poses a
riddle:
On the one hand it strongly points toward photoionization by a central
power-law source as the nebula's excitation mechanism (\S~\ref{sec41}),
yet on the other hand the nucleus of \n7252 appears devoid of any such
power-law source and seems, therefore, to be presently inactive
(\S~\ref{sec42}).

One possible solution is that the \oiiineb\ could---like the by now
well-studied \oiii-luminous nebula nicknamed {\em Hanny's Voorwerp}\,
near IC~2497 \citep{lint09,jozs09,ramp10,scha10,keel12b}---be an
ionization echo revealing relatively recent ($\sim$\,10$^4$\,--\,10$^5$~yr),
but now declined AGN activity in its host galaxy.
If so, how long past the cessation of central AGN activity could the
nebula keep glowing and keep its high-excitation signatures?

For a nebula consisting mostly of hydrogen, the recombination time is
$\trec = [\reccoHtwo\,\nel]^{-1}\approx 10^5/\nel$ years for
$T\approx 10^4$ K, where \reccoHtwo\ is the total recombination coefficient
for Case B \citep[esp.\ Table~2.1]{of06} and \nel\ is the local electron
density.
In the case of the \oiiineb, $T\la 8900$~K, $\reccoHtwo\ga 2.83\times
10^{-13}$ cm$^3$ s$^{-1}$, and $\nel\approx 10$\,--\,100 cm$^{-1}$, whence
the hydrogen recombination time is $\trec\approx 1100$\,--\,11,300 yr.
Clearly, if its source of excitation were suddenly turned off the nebula
would fade on a short time scale ($\sim\!10^4$ yr).
Since different ionic species differ in their \trec, the observed line
ratios would change as well \citep{bine87}.
Specifically, the excitation-sensitive ratios $\oiiisev/\hbet$ and
$\neiiieig/\hbet$ would decline rapidly, since the recombination times of
both O$^{++}$ and Ne$^{++}$ are an order of magnitude shorter than
\trec(H$^+$).
Hence, the \oiii\ luminosity of the nebula would begin fading on a very short
time scale, likely $\la\!10^3$ yr.
It would, though, take the fading about 20,000\,--\,50,000 years to progress
through the nebula because of its $\sim$\,$7\times 10$ kpc projected size
(\S~\ref{sec21}) and the finite speed of light.

Yet, the value measured for the excitation of the \oiiineb,
$\oiiisev/\hbet = 10.03 \pm 0.25$ (Table~\ref{tab03}), is close to the maximum values
observed in Seyfert galaxies and EELRs around quasars (Fig.~\ref{fig05}).
Hence, the nebula was clearly {\em still being excited,} most likely by a
source of strongly ionizing radiation (\S~\ref{sec41}), when it emitted the
photons recorded in our images and spectra.
As \citet{bine87} concluded already long ago for similar \oiii-bright
emission-line regions around galaxies with AGNs, this rules out any ``fossil
nebula'' interpretation.
What, then, is the minimum luminosity of the ionizing source required to
ionize the nebula if the source lies in the nucleus, and what is the light
travel time from the nucleus to the nebula, during which the nucleus could
have faded?

A lower limit to the luminosity of the central ionizing source can be
estimated from the recombination lines of the \oiiineb\ and, specifically,
from the observed \hbet\ flux.
As \citet{zans27} first pointed out for a pure hydrogen nebula excited by
a central star, the number of \hbet\ photons emitted by the nebula is related
to the number of ionizing photons from the central source (star) through the
relation
$$
  \left(\frac{L_{{\rm H}\beta}}{h\nu_{{\rm H}\beta}}\right)_{\!\!\rm neb}
  \approx \frac{\reccohbeteff}{\reccoHB} \cdot \omega \cdot
  \!\int_{\nu_1}^{\nu_2}\!\left(\frac{L_{\nu}}{h\nu}\right)_{\!\!\rm s}d\nu ,
$$
where \reccohbeteff\ is the effective recombination coefficient for \hbet\
(Case B), \reccoHB\ the total hydrogen recombination coefficient for Case B,
and $\omega$ the covering factor of the nebula as seen from the central
source.
In a real nebula also containing helium, the main wavelength range of UV
photons capable of ionizing hydrogen lies between the ionization limit of
helium at 228\,\AA\ (54.4 eV) and that of hydrogen at 912\,\AA\ (13.6 eV).
As \citet{keel12a} point out for the case of a nebula excited by an AGN, the
luminosity of the central source must be sufficient to explain all regions
of \hbet\ emission, including those of {\em highest\,} surface brightness;
in the present case of the \oiiineb, the peak surface brightness measured
in the MagE slit corresponds to $F_0(\hbet) = 1.7\times 10^{-17}$ \ergcms\ 
per square arcsecond.
Using their formula $\lion = 1.3\times 10^{64} z^2 F_0(\hbet)/\alpha^2$,
where $z$ is the redshift and $\alpha$ the angle of the area under
consideration (here $1\arcsec\times 1\arcsec$) as seen from the central
source and expressed in degrees, yields a minimum ionizing luminosity of
$\lion \ga 5\times10^{42}$ \ergsec\ for the central source in \n7252
as seen by the \oiiineb.

For an ionizing continuum typical of AGNs, $L_{\nu} \propto \nu^{\alpha}$
with $\alpha \approx -1.0$, the energy $\nu L_{\nu}$ is approximately
constant per decade, whence the minimum X-ray luminosity should be
comparable to the above-derived minimum ionizing luminosity.
Yet, our estimated upper limit for the current nuclear X-ray luminosity
of \n7252 is $\lxttzero < 5\times 10^{39}$ \ergsec\ (\S~\ref{sec42}), or
fully three orders of magnitude lower than expected.
This gross mismatch suggests that the nucleus experienced strongly increased
activity until recently, sufficient to photoionize the \oiiineb\ as observed,
yet this activity has declined since then dramatically.

The light-travel time from the nucleus to the \oiiineb\ could be as short
as $\sim$\,18,000~yr if the nebula were to lie at a true distance equal to
its 5.4~kpc projected distance (for its brightest part) from the nucleus,
i.e., in the same sky plane as the nucleus, or it could be even shorter
if the nebula were to lie closer to us than the nucleus, i.e., in front of
that sky plane.
However, the kinematics of the nebula suggests that it is part of the tidal
stream of infalling gas (\S~\ref{sec34}), which two best-fit models place
mostly {\em behind\,} that plane \citep{himi95,chba10}.
Unfortunately, these models do not reproduce the \hi\ observations in
sufficient detail to permit locating the nebula accurately along our
line of sight (J.\ E.\ Barnes, private communication).
They only suggest that as part of the tidal-gas stream the nebula must
lie somewhere between the sky plane and $\sim\,$30~kpc behind it.
This leaves us to guess that the increase in light-travel time from the
nucleus to us via the \oiiineb\ is of the order of 20,000\,--\,200,000~yr.

We conclude that (1) the \oiiineb\ appears to be the ionization echo of some
AGN activity in \n7252 that lasted until about 20,000\,--\,200,000 yr ago,
and (2) this nuclear activity has declined by at least three orders of
magnitude since then.
Note that the decline could easily exceed three orders of magnitude since
the minimum ionizing luminosity derived above ($\ga 5\times10^{42}$ \ergsec)
is based on the projected distance of the nebula from the nucleus (5.4 kpc),
while the true distance may be up to $\sim\,$6 times larger.
Perhaps the single most important assumption made in the above arguments is
that the nuclear source emits isotropically.
Were the emission to be beamed toward the \oiiineb\ instead, the inferred
decline of nuclear activity could be significantly less.
We comment on this and other alternative interpretations further in
\S~\ref{sec46} below.

\vskip 0.5truecm
\subsection{Comparison with Hanny's Voorwerp and Similar EELRs}
\label{sec44}

How does the \oiiineb\ of \n7252 compare with the prototypical `Hanny's
Voorwerp' of IC~2497 and with other extended emission-line regions also
thought to be ionization echoes?

Table~\ref{tab05} summarizes the main parameters derived for the \oiiineb\ in the
present paper and for Hanny's Voorwerp by \citet{lint09}, \citet{scha10},
and \cite{keel12b}.
As a comparison of the individual parameters shows, the two nebulae are
surprisingly similar in many of their properties.
The main difference is that the \oiiineb\ appears to be a scaled down version
of Hanny's Voorwerp, being about 3$\times$ smaller in linear projected size
and 100$\times$ less luminous in \hbet\ and \oiiisev.
Note also that its metallicity is about twice that of Hanny's Voorwerp,
though still clearly subsolar, and its electron temperature is significantly
lower.
This lower temperature only strengthens the case that, like Hanny's Voorwerp,
the \oiiineb\ is unlikely to be shock excited (\S~\ref{sec41}) and must,
therefore, be photoionized.

EELRs were first discovered around QSOs and are recognized as \oiii-bright
emission nebulosities extending well beyond the limit of central narrow-line
regions ($r\la 1$~kpc), typically to distances of 10\,--\,40 kpc from the
center.
They are often, but not always, associated with gaseous tidal debris being
photoionized by a central engine \citep{stoc83,stoc87,stoc06}.
Their typical projected size is $\sim\,$10 kpc \citep{huse13}, in accord
with that of the \oiiineb.
A somewhat arbitrary minimum line luminosity of
$L_{\rm 0,min}(\oiiisev) = 5\times 10^{41}$ \ergsec\ is often associated
with the definition of an EELR \citep[e.g.,][]{fust09,huse13}.
According to this restrictive definition, Hanny's Voorwerp clearly is an EELR,
while \n7252's \oiiineb\ is not, falling short by a factor of $\sim\,$36 in
$L_0(\lambda 5007)$.
Yet, the two nebulae are so similar in most of their properties, including
their being associated with gaseous tidal debris, that we believe it more
logical to regard \n7252's nebula as an EELR of very low luminosity.

A comparison of the nuclear ionizing luminosities and decline times inferred
from the \oiii\ nebula and Hanny's Voorwerp for \n7252 and IC~2497, respectively,
is also of interest.
As Table~\ref{tab05} shows, the minimum nuclear \lion\ for \n7252 is three orders of
magnitude lower than that for IC~2497, suggesting that the nuclear activity
still seen in \n7252 by the \oiiineb\ is much less than that seen by Hanny's
Voorwerp in IC~2497.
In both galaxies, the present X-ray luminosity in the 2\,--\,10 keV band is
at least three orders of magnitude lower than the ionizing luminosity still
exciting the EELR,\footnote{
Neither \n7252 nor IC~2497 was detected in the 14\,--\,195 keV band by the
{\em Swift}/BAT All-Sky Hard X-Ray Survey \citep{baum13}, thus eliminating
the possibility of highly obscured strong AGN activity in both galaxies.}
and the inferred decline times of about $2\times 10^4$\,--\,$2\times 10^5$ yr
are similar.
However, note that the minimum ionizing luminosity inferred for IC~2497,
$\lion\ga (2$\,--\,$7)\times 10^{45}$ \ergsec\ \citep{scha10,keel12b} clearly
attributes a recent QSO-like luminosity ($\geq 10^{44}$ \ergsec) to the
galaxy's nucleus, whereas \n7252's $\lion\ga 5\times 10^{42}$ \ergsec\ only
attributes a recent {\em Seyfert\,}-like luminosity (10$^{42}$\,--\,10$^{44}$
\ergsec) to this merger remnant's nucleus.

Studying a sample of 19 galaxies with EELRs at projected radii $>$10 kpc,
\citet{keel12a} have recently identified seven additional cases where the
EELR is likely to be an ionization echo revealing {\em past\,} enhanced AGN
activity in the host galaxy.
The seven host galaxies have redshifts in the range of $z = 0.022$\,--\,0.091
and inferred minimum ionizing luminosities of
$\lionmin = 8\times 10^{43}$\,--\,$1.3\times 10^{45}$ \ergsec\ necessary to
excite their EELRs.
Hence, if we add \n7252 and IC~2497 to this sample of candidate systems with
ionization echos, \n7252 with its \oiiineb\ clearly marks the low-energy
end of the sequence, while IC~2497 with its Voorwerp marks the high-energy
end.
At $z = 0.0158$, \n7252 is the nearest of the nine objects, but not by much.
Hence, the fact that its $\lionmin = 5\times 10^{42}$ \ergsec\ is more than
an order of magnitude lower than that of the next member of the sequence
($8\times 10^{43}$ \ergsec) makes \n7252 a host galaxy with a low-luminosity
ionization echo and exceptionally faded AGN activity.
In contrast, IC~2497 and Hanny's Voorwerp at $z = 0.0499$ represent a
relatively nearby case of recent QSO-like luminosity followed by a very
strong decline, explaining why this system was discovered first and will
likely remain the prototype of its class.

In short, \n7252's \oiiineb\ seems to be the lowest-luminosity ionization
echo discovered so far.
The nucleus of \n7252 itself must have reached at least Seyfert-like
luminosity to still excite the nebula as observed, yet has since faded by
several orders of magnitude to a very low, so far undetected activity level.
The fact that the \oiiineb\ is clearly associated with gaseous tidal debris
adds to the growing evidence that many ionization echos occur in just such
debris \citep{stoc83,jozs09,keel12a}.

\subsection{Implications for the History of \n7252}
\label{sec45}

It is interesting to place the above results on the \oiiineb\ and inferred
nuclear activity into the context of \n7252's merger history.

According to the best currently available model simulation (\citealt{chba10};
see also \citealt{himi95}), this major merger began with the first close
approach of two gas-rich disk galaxies about 640 Myr ago and was essentially
complete when the two nuclei coalesced $\sim\,$220 Myr ago (times scaled to
$H_0 = 73$ \kms\ Mpc$^{-1}$).
During this period, shock-induced star and cluster formation was enhanced by
a factor of $\sim\,$4 over normal, leading to the spectacular set of
several hundred young halo globular clusters \citep{mill97,ss98}, and peaking
shortly after the coalescence of the two nuclei (see Fig.~4 of
\citealt{chba10}).
Presently, about 220 Myr after coalescence, intense star and cluster
formation continues mainly in the central gas disk (``minispiral,''
\S~\ref{sec31}).

If our conclusion is correct that over the past 0.02\,--\,0.20 Myr the
nuclear activity has declined from a Seyfert-like level by at least three
orders of magnitude (\S~\ref{sec44}), then it must have fluctuated by
comparable amounts over the past 200$^+$\,Myr.
Otherwise, we would be privileged to observe \n7252 just at the moment when
its previous, approximately steady Seyfert-like activity suddenly declined;
this seems highly unlikely.
Hence, we conclude that {\em the \oiiineb\ must be witness to some sputtering
late AGN activity with surprisingly large luminosity amplitudes ($\ga 10^3$)
in \n7252.}

Yet, even at the present time the central gas disk contains
$\sim$\,$5\times 10^9$ \msun\ of molecular gas \citep{dupr90,wang92}, and
the tidal tails and loop contain another $\sim$\,$3.8\times 10^9$ \msun\ of
\hi\ \citep{hibb94}.
Hence, {\em if\,} the nucleus reached much higher, QSO-like levels of activity 
in the past---as might be expected around or shortly after the time of
nuclear coalescence---then these hypothetical high levels of nuclear activity
have failed to rid the remnant of its gas, both in the central disk and in
the tidal tails and W Loop.
This sobering fact should temper any theoretical assumptions that gas-rich
major mergers and their ensuing nuclear activity necessarily clean the
remnant galaxies to a ``red-and-dead'' state.

\subsection{Remaining Questions and Suggested Further Work}
\label{sec46}

There are some obvious questions concerning the \oiiineb\ that we haven't
addressed so far.
Perhaps most important among them are:
(1) Why does the \oiiineb\ extend only over $\sim\,$10 kpc in projection
if really it forms part of a much longer stream of tidal-tail gas falling
back into the merger remnant?
And (2), what causes the striking filaments (``streamers,'' Fig.~\ref{fig01}
and \S~\ref{sec31}) in the \oiiineb\ that seem to extend from the bright
ridge of \oiiisev\ emission to the SSE?

In answer to the first question, it is tempting to speculate that ionizing
radiation may escape---or have escaped---from the nucleus of \n7252 in a
cone-shaped volume, perhaps constrained by some central torus of dense
circum-nuclear gas, as is often the case in Seyfert galaxies.
If so, the \oiiineb\ simply marks the intersection of this cone-shaped
volume with the eastern gaseous tidal tail.

Alternative explanations cannot, however, be clearly ruled out.
A detailed comparison of the \oiii\ and \halp\ images of the \oiiineb\
(Fig.~\ref{fig01}{\em c}\,--\,{\em f}\,) with Fig.~7 of \citet{hibb94} shows
that the westernmost ``streamer'' of the \oiiineb\ coincides closely with a
steep drop in the \hi\ surface mass density, marked also by a strong \hi\
velocity gradient.
It is conceivable that a so far undetected plasma jet from the nucleus is
ablating the \hi, which might explain the limited extent of the nebula, or
that an already more centrally-wrapped stream of gas is slamming into the
tidal-tail \hi\ stream and shock-ionizing it.
Our strongest arguments against both possibilities have been provided by
(i) the narrow velocity widths of most nebular emission lines
(\S~\ref{sec331}), (ii) the low temperature of the nebula (\S~\ref{sec41}),
and (iii) the apparent continuity in line-of-sight velocities along the
tidal-tail stream and nebula (\S~\ref{sec34}, esp.\ Fig.~\ref{fig06}).
Yet, the presence  of some unexplained broader line wings near the base of
especially the \niiboth\ lines (Fig.~\ref{fig04}{\em c}) may indicate
increased turbulence in at least part of the gas of the nebula.
Spectral mapping of the entire \oiiineb\ with an IFU spectrograph at a high
signal-to-noise ratio (much higher than that of our MagE spectrum) will be
required to address this issue.

Such mapping might also help address our question about the nature and
origin of the striking {\em filaments}.
It would yield valuable quantitative information concerning the density
and excitation variations present across the \oiiineb\ and filaments.
A visual intercomparison of Figs.~\ref{fig01}{\em e} and \ref{fig01}{\em f\,}
suggests that the excitation ($\oiiisev/\hbet$) of the easternmost filament
is significantly lower than that of the westernmost filament
(\S~\ref{sec31}), a fact remaining to be explained.
Perhaps the most interesting feature of the filaments is their projected
geometry: they seem to run more nearly {\em perpendicular} to any radius
drawn from the nucleus than parallel (Fig.~\ref{fig01}{\em c}\,--\,{\em f}).
This geometry would seem to argue against the filaments being minor gas
streams flowing away from the nucleus (driven, e.g., by a hypothetical jet)
or falling toward it.
Rather, they may represent either density variations in the gas
or---perhaps---fine structure in the ionization echo
indicative of temporal variations in the central ionizing radiation.
The above-mentioned spectral mapping seems then to be a necessary first step
toward a better understanding of their nature and origin.

A third interesting question is whether there is any likely connection
between the presence of the \oiiineb\ and the presence of the string of
\hii\ regions further downstream in the W Loop (Fig.~\ref{fig06} and
\S\S~\ref{sec31}, \ref{sec34}).
Such a connection could, e.g., be created by any mechanism that currently
excites the \oiiineb\ and whose direction remains fixed over a prolonged
period.
If so, could the same mechanism have compressed the gas from which the
string of \hii\ regions has formed?
The answer is very likely negative, as the following simple estimate shows.
The projected distance from the center of the \oiiineb\ to the approximate
center of the five \hii\ regions is about $25\arcsec\approx 8.0$ kpc.
The tidal-tail gas near \hii(S101) moves with $\dvlos = -230\pm 5$ \kms\
(\S~\ref{sec34}) which, according to the best model simulations of \n7252
\citep{himi95,chba10}, must be close to the fallback velocity within the
E tail.
Traversing the above projected distance at this velocity takes about 35 Myr,
or easily 5\,--\,10$\times$ the age of a giant \hii\ region excited by
O stars.
Since the true distance between the \oiiineb\ and the \hii\ regions is
probably considerably larger than the projected 8 kpc, the collapse of the
tail gas hypothetically triggered by exposure to the same mechanism that
presently excites the \oiiineb\ would have taken an order of magnitude longer
than the age of the \hii\ regions.
This seems unlikely, though not necessarily impossible.

Alternatively, however, either a directional change of some hypothetical jet
or a past more isotropic event could have---in principle---both cleared the
path to the \oiiineb\ and caused the formation of the \hii\ regions by
compressing the gas there and inducing star formation.
Such complex phenomena are observed in, e.g., the NE radio lobe of \n5128
\citep{blan75,grah81,morg91,oost05}.

In order to better understand these various issues, {\em improved model
simulations} of the merger that occurred in \n7252 will clearly be at least
as important as new observations.
Figuring out the true spatial geometry of the ionized gas in this nearby
merger remnant holds great promise, since it will help pin down the time
delays involved in the ionization history.
However, it will take major new modeling efforts with improved gas
hydrodynamics to fully realize this promise.

\section{SUMMARY}
\label{sec5}

We have described imaging and spectroscopic observations of a newly
discovered extended emission-line nebulosity in \n7252, dubbed the
`\oiiineb.'
The observations, made at Las Campanas Observatory, consist of broad-
and narrow-band images obtained with the du Pont 2.5 m telescope and
follow-up long-slit spectra obtained with the LDSS-3 and MagE spectrographs
at the Clay 6.5 m telescope.
The discovery of this $\sim\,$10 kpc-sized nebula is of interest because
it seems to yield the first sign ever of episodic AGN activity still
occurring in this prototypical merger remnant, about 220 Myr after the
coalescence of the two nuclei of the participant spiral galaxies.

The main conclusions from our observations and analysis concerning this
\oiiineb\ and the remnant's nuclear activity are as follows.

1. The \oiiineb\ lies about 4 kpc west and 3.6 kpc south (brightest part)
of the nucleus and covers a projected area of $10.1\times 6.6$ kpc$^2$
($32\arcsec\times 21\arcsec$) in the E-W and N-S directions.
Its main bright ridge appears approximately aligned with the E tidal tail,
which is known to pass behind the main body of the remnant, and the
southern leg of the W Loop.
Several long faint filaments emerge from this ridge in the SSE direction.
Line-of-sight velocities measured across the nebula show a smooth velocity
gradient that seems to fit in with \hi\ velocities in the E tail and W Loop,
suggesting that the nebula belongs to the stream of tidal-tail material
falling back into the remnant.

2. The UV--optical spectrum of the bright ridge recorded at a resolution
of $R\approx 4100$ with the MagE spectrograph shows---besides the dominant
\oiii\ and \oii\ doublets---the Balmer recombination lines \halp\,--\,\hdel,
the usual collisionally-excited doublets of \nii\ and \sii, and the
relatively high-excitation lines of \neiiieig\ and \heiisix.
The latter requires full ionization of helium and, hence, an ionization
source rich in photons of energy $h\nu > 54.4$ eV.
The lines are generally narrow, with a FWHM of $22.7\pm 3.3$ \kms\ that
corresponds to a mean velocity dispersion in the ionized gas of
$\sigvelgas = 9.6\pm 1.4$ \kms, or roughly the sound speed at $T = 8000$\,K.

3. The mean electron density, temperature, and internal extinction of the
nebula are all low.
We measure an electron density of\, $\nel \la (2$\,--\,$3)\times 10^2$
cm$^{-3}$---with likely values mostly $\la\,$10 cm$^{-3}$---in the bright
ridge covered by the MagE slit, and find an upper limit of
$\nel < 500$ cm$^{-3}$ along the entire $\sim\,$10 kpc slice of the nebula
covered by the LDSS-3 slit.
In agreement with these low densities, the internal visual extinction
determined from the Balmer-line decrement is also low, $A_V = 0.15\pm 0.15$
mag.
Given the high nebular excitation inferred from the flux ratio
$F_0(\oiiisev)/F_0(\hbet) = 10.03\pm 0.25$, our upper limit for the electron
temperature, $T < 8900$\,K, is surprisingly low.
Finally, at the distance of 66.2 Mpc adopted for \n7252 the integrated
luminosity of the nebula is $L_0(\lambda 5007) = (1.37\pm 0.25)\times 10^{40}$
\ergsec\ in the strongest \oiii\ line and $L_0(\halp) = (4.2\pm 0.9)\times
10^{39}$ \ergsec\ in the strongest Balmer line.

4. In three standard diagnostic diagrams, the line ratios measured for the
\oiiineb\ place it clearly in the domain of Seyfert galaxies and well away
from the dividing line between Seyferts and LINERS.
Several arguments suggest that {\em the nebula is most likely photoionized
by a central source in \n7252 with a power-law spectrum.}
The alternative possibility that the nebula might be excited by high-velocity
shocks of $\vshock \ga 450$ \kms, necessary to explain its measured high
excitation, seems unlikely because of the low nebular temperature
($<\,$8900\,K) and internal velocity dispersion ($\sim\,$9.6 \kms).
A comparison with photoionization models in the literature then suggests a
subsolar gas metallicity of $\sim$\,$2/3$ \zsun.
This rough estimate is consistent with the metallicity of $Z = 0.80\pm 0.07$
\zsun\ that we determine for \hiisone, a 30 Doradus-like giant \hii\ region
located 8 kpc in projection downstream from the nebula.
Our conclusion that the \oiiineb\ is photoionized by a central power-law
source in \n7252 is further supported by the fact that in the
above-mentioned diagnostic diagrams it falls among a group of EELRs observed
by \citet{fust09} around five quasars, where photoionization is the only
viable explanation.

5. Yet, an observational and literature search for AGN activity in
\n7252---from X-rays through the UV--optical to near-IR, mid-IR, and radio
wavelengths---fails to yield any significant detection.
Although vigorous star formation occurs in the central gas disk of
$\sim\,$1.6 kpc radius at the rate of $\sim\,$6.3 \msunyr, the nucleus itself
appears presently inactive.
The perhaps most stringent upper limit for any AGN activity is set by X-ray
observations of \n7252, from which we estimate an upper luminosity limit of
$\lxttzero < 5\times 10^{39}$ \ergsec\ for the nucleus.
This upper limit is three orders of magnitude lower than the minimum nuclear
ionizing luminosity of $\lion \ga 5\times 10^{42}$ \ergsec\ necessary to
excite the \oiiineb\ and produce its observed \hbet\ emission.
Hence, there is a discrepancy of at least $10^3$ between the ionizing
luminosity that the nebula was exposed to when it emitted the photons
recorded on our images and spectra, and the low luminosity of the presently
observed nucleus.

6. Assessing all available evidence, we conclude that the \oiiineb\ of n7252
is most likely a {\em faint ionization echo.}
It shows clear signs of being excited by a mildly active nucleus, yet the
ionizing luminosity of this nucleus has declined by at least three orders of
magnitude over the past 20,000\,--\,200,000 years.
In many ways this nebula resembles the prototypical `Hanny's Voorwerp,' a
bright likely ionization echo recently discovered near the still weakly
active galaxy IC~2497; but in \n7252 its size is 3$\times$ smaller and its
\oiii\ luminosity $\sim\,$100$\times$ lower.
We propose that it should be classified as an EELR (extended emission-line
region) despite falling well short of the minimum line luminosity of
$L(\lambda 5007) \geq 5\times 10^{41}$ \ergsec\ often adopted in
the definition of EELRs.
If so, the \oiiineb\ is the lowest-luminosity ionization echo and EELR
discovered so far, indicative of recent, probably {\em sputtering AGN activity
of Seyfert-like intensity in \n7252}.
It is the first sign of any AGN activity found in this merger remnant,
occurring about 640 Myr after the merger of two gas-rich disk galaxies began
and $\sim\,$220 Myr after the two nuclei coalesced.
Its proximity and the rich multi-wavelength data---including detailed gas
kinematics---available for \n7252 hold great promise for further study of
black-hole feeding and feedback.
This promise should encourage new model simulations---with improved
hydrodynamics---of this intriguing merger remnant and its gaseous contents.

\acknowledgments

We thank
Herman Olivares, Jorge Araya, Hern\'an Nu\~nez, and Mauricio Martinez for
expert assistance at the telescopes;
Joshua Barnes for helpful advice and correspondence on model simulations of
\n7252;
Hai Fu for data on EELRs and help with producing Fig.~\ref{fig05};
John Hibbard for permission to reproduce his \hi\ maps in Fig.~\ref{fig07};
and Jack Baldwin, Vardha N.\ Bennert, Jeremy Darling, Luis Ho, Barry
Madore, Eric Murphy, Ian Thompson, and Sylvain Veilleux for helpful
information and discussions. 
Figures~\ref{fig01} and \ref{fig02} were prepared with the astronomical
graphics package WIP \citep{morg95}, maintained and generously made available
by James A.\ Morgan and Peter Teuben.
This research has greatly benefited from the use of
(1) NASA's Astrophysics Data System (ADS) Bibliographic Services, operated
by the Smithsonian Astrophysical Observatory, and
(2) the NASA/IPAC Extragalactic Database (NED), which is operated by the
Jet Propulsion Laboratory, California Institute of Technology, under
contract with NASA.
One of us (F.S.) acknowledges partial, but crucial support from the NSF
through grant AST-99\,00742.


%
%
%
%

\clearpage

\begin{deluxetable*}{llllcrrccl}
\tablecolumns{10}
\tablewidth{0pt}
\tablecaption{Log of Observations of NGC 7252\label{tab01}}
\tablehead{
  \colhead{}             &
  \colhead{}             &
  \colhead{}             &
  \colhead{}             &
  \colhead{}             &
  \colhead{}             &
  \colhead{Total\ }      &
  \colhead{Wavelength}   &
                           \\
  \colhead{}             &
  \colhead{}             &
  \colhead{}             &
  \colhead{CCD}          &
  \colhead{}             &
  \colhead{P.A.}         &
  \colhead{\,Expos.}     &
  \colhead{Coverage}     &
  \colhead{Seeing}       &
                           \\
  \colhead{Date}         &
  \colhead{Telescope}    &
  \colhead{Instrument\tablenotemark{a}} &
  \colhead{Detector}     &
  \colhead{Filter}       &
  \colhead{(deg)}        &
  \colhead{(s)}          &
  \colhead{(\AA)}        &
  \colhead{(arcsec)}     &
  \colhead{Notes\ \ }
}
\startdata
2000 Sep 27 ....&  du Pont 2.5 m& DC    & Tek \#5 & [O\,III]&  1.2& 15300& 5071--5102& 0.9& \nodata  \\
                &               &       &         & BluCont &  1.2&  2520& 5180--5495& 1.0& \nodata  \\
2000 Sep 28 ....&  du Pont 2.5 m& DC    & Tek \#5 & H$\alpha$& 1.2& 12300& 6651--6702& 0.8& \nodata  \\
                &               &       &         & RedCont &  1.2&   900& 6450--7145& 0.7& \nodata  \\
2000 Sep 29 ....&  du Pont 2.5 m& DC    & Tek \#5 & $B$     &  1.2&  4800& 3850--4850& 0.7& \nodata  \\
                &               &       &         & $V$     &  1.2&  2520& 4900--5800& 0.8& \nodata  \\
2000 Sep 30 ....&  du Pont 2.5 m& DC    & Tek \#5 & [S\,II] &  1.2&  5400& 6787--6865& 0.7& \nodata  \\
                &               &       &         & RedCont &  1.2&   900& 6450--7145& 0.6& \nodata  \\
2007 Sep 5 ......& Clay 6.5 m   & LDSS-3& STA0500A& \nodata& 103.3&  7200& 3850--6590& 0.7& Slit $1\farcs1\times 70\arcsec$\\
2007 Sep 5--6 ..&  Clay 6.5 m   & LDSS-3& STA0500A& OG\,590& 103.3& 12600& 6070--10760&0.5--0.8& Slit $1\farcs1\times 70\arcsec$\\
2007 Sep 6 ......& Clay 6.5 m   & LDSS-3& STA0500A& \nodata& 103.3&  7200& 3850--6590& 0.6--1.0& Slits $0\farcs9\times 70\arcsec$ (4)\\
2009 Aug 23 ... &  Clay 6.5 m   & MagE  & E2V 42-20&\nodata& 103.3&  9000& 3300--8250& 0.7& Slit $1\farcs0\times 10\arcsec$
\enddata
\tablenotetext{a}{DC = Direct camera;\ \ LDSS-3 = Low-Dispersion Survey Spectrograph 3;\ \ MagE =
Magellan Echellette Spectrograph.}
\end{deluxetable*}

\begin{deluxetable*}{lccccc}
\tablecolumns{6}
\tablewidth{0pt}
\tablecaption{Emission-Line Fluxes of \hii\,(S101)\label{tab02}}
\tablehead{
   \colhead{}          &
   \colhead{$\lambda$} &
   \colhead{Flux $F$\tablenotemark{a}}  &
   \colhead{}          &
   \colhead{}          &
   \colhead{}              \\
   \colhead{\phn Line/Ion\phn}  &
   \colhead{(\AA)}     &
   \colhead{($10^{-16}$ erg cm$^{-2}$ s$^{-1}$)} &
   \colhead{\lineratzeromw\tablenotemark{b}}  &
   \colhead{}          &
   \colhead{\lineratzero\tablenotemark{c}}
}
\startdata
\hdel\dotfill   &   4101 &   $0.51\pm 0.03$ &    $0.212\pm 0.013$ &&   $0.253\pm 0.015$  \\
\hgam\dotfill   &   4340 &   $1.21\pm 0.04$ &    $0.500\pm 0.017$ &&   $0.569\pm 0.020$  \\
\oiii\dotfill   &   4363 &      $\la 0.08$  &        $\la 0.033$  &&       $\la 0.036$   \\
\hbet\tablenotemark{d}\dotfill &
                    4861 &   $2.46\pm 0.03$ &    $1.000\pm 0.000$ &&   $1.000\pm 0.000$  \\
\oiii\dotfill   &   4959 &   $0.72\pm 0.01$ &    $0.292\pm 0.007$ &&   $0.285\pm 0.007$  \\
\oiii\dotfill   &   5007 &   $2.31\pm 0.03$ &    $0.937\pm 0.017$ &&   $0.902\pm 0.016$  \\
\hei\dotfill    &   5876 &   $0.27\pm 0.01$ &    $0.106\pm 0.004$ &&   $0.086\pm 0.003$  \\
\oi\dotfill     &   6300 &   $0.24\pm 0.01$ &    $0.094\pm 0.004$ &&   $0.071\pm 0.003$  \\
\nii\dotfill    &   6548 &   $1.20\pm 0.03$ &    $0.473\pm 0.013$ &&   $0.349\pm 0.010$  \\
\halp\tablenotemark{e}\dotfill &
                    6563 &  $10.24\pm 0.10\phn$& $4.030\pm 0.061$ &&   $2.965\pm 0.046$  \\
\nii\dotfill    &   6583 &   $3.42\pm 0.02$ &    $1.347\pm 0.018$ &&   $0.989\pm 0.014$  \\
\sii\dotfill    &   6716 &   $1.83\pm 0.02$ &    $0.720\pm 0.012$ &&   $0.521\pm 0.008$  \\
\sii\dotfill    &   6731 &   $1.26\pm 0.02$ &    $0.496\pm 0.010$ &&   $0.358\pm 0.007$
\enddata
\tablenotetext{a}{Measured through aperture of $4\farcs72\times 1\farcs1$ centered on
		  Cluster S101, corresponding to a projected area of $1.51\times 0.35$
		  kpc$^2$ at the distance of \n7252.}
\tablenotetext{b}{Ratio of fluxes corrected for Milky Way foreground extinction of
		  $\avmw = 0.10$.}
\tablenotetext{c}{Ratio of fluxes corrected for total extinction (Milky Way plus
		  internal) of $\av = 1.0 \pm 0.2$.}
\tablenotetext{d}{Reddening-corrected flux within aperture is\ \ $F_0({\rm H}\beta) =
		  (7.3\pm 1.6)\times 10^{-16}$ erg cm$^{-2}$ s$^{-1}$.}
\tablenotetext{e}{Reddening-corrected flux within aperture is\ \ $F_0({\rm H}\alpha) =
		  (21.6\pm 3.2)\times 10^{-16}$ erg cm$^{-2}$ s$^{-1}$, with
		  corresponding luminosity of\ \ $L_0({\rm H}\alpha) = (1.13\pm 0.17)
		  \times 10^{39}$ erg s$^{-1}$.}
\end{deluxetable*}

\begin{deluxetable*}{lccccc}
\tablecolumns{6}
\tablewidth{0pt}
\tablecaption{Emission-Line Fluxes of \oiii\ Nebula\label{tab03}}
\tablehead{
   \colhead{}          &
   \colhead{$\lambda$} &
   \colhead{Flux $F$\tablenotemark{a}}  &
   \colhead{}          &
   \colhead{}          &
   \colhead{}              \\
   \colhead{\phn Line/Ion\phn}  &
   \colhead{(\AA)}     &
   \colhead{($10^{-16}$ erg cm$^{-2}$ s$^{-1}$)} &
   \colhead{\lineratzeromw\tablenotemark{b}}  &
   \colhead{}          &
   \colhead{\lineratzero\tablenotemark{c}}
}
\startdata
\nev\dotfill    &   3426 &       $<0.20$    &        $<0.275$    &&        $<0.293$     \\
\oii\dotfill    &   3726 &   $2.27\pm 0.06$ &   $3.087\pm 0.082$ &&   $3.229\pm 0.145$  \\
\oii\dotfill    &   3729 &   $3.51\pm 0.09$ &   $4.773\pm 0.122$ &&   $4.992\pm 0.224$  \\
\neiii\dotfill  &   3868 &   $0.88\pm 0.03$ &   $1.198\pm 0.040$ &&   $1.245\pm 0.048$  \\
\hdel\dotfill   &   4101 &   $0.18\pm 0.01$ &   $0.240\pm 0.013$ &&   $0.247\pm 0.014$  \\
\hgam\dotfill   &   4340 &   $0.29\pm 0.02$ &   $0.385\pm 0.027$ &&   $0.394\pm 0.028$  \\
\oiii\dotfill   &   4363 &       $<0.03$    &        $<0.040$    &&        $<0.041$     \\
\heii\dotfill   &   4686 &   $0.24\pm 0.02$ &   $0.320\pm 0.020$ &&   $0.322\pm 0.020$  \\
\hbet\tablenotemark{d}\dotfill &
                    4861 &   $0.76\pm 0.02$ &   $1.000\pm 0.026$ &&   $1.000\pm 0.026$  \\
\oiii\dotfill   &   4959 &   $2.45\pm 0.06$ &   $3.223\pm 0.080$ &&   $3.209\pm 0.080$  \\
\oiii\dotfill   &   5007 &   $7.68\pm 0.19$ &  $10.089\pm 0.249\phn$&& $10.026\pm 0.247\phn$ \\
\nii\dotfill    &   5755 &       $<0.02$    &        $<0.026$    &&        $<0.025$     \\
\hei\dotfill    &   5876 &       $<0.03$    &        $<0.039$    &&        $<0.038$     \\
\oi\dotfill     &   6300 &   $0.32\pm 0.02$ &   $0.413\pm 0.026$ &&   $0.394\pm 0.025$  \\
\nii\dotfill    &   6548 &   $0.40\pm 0.02$ &   $0.508\pm 0.026$ &&   $0.483\pm 0.025$  \\
\halp\tablenotemark{e}\dotfill &
                    6563 &   $2.54\pm 0.08$ &   $3.242\pm 0.102$ &&   $3.081\pm 0.157$  \\
\nii\dotfill    &   6583 &   $1.38\pm 0.08$ &   $1.761\pm 0.102$ &&   $1.673\pm 0.097$  \\
\sii\dotfill    &   6716 &   $0.70\pm 0.02$ &   $0.895\pm 0.025$ &&   $0.848\pm 0.046$  \\
\sii\dotfill    &   6731 &   $0.53\pm 0.03$ &   $0.680\pm 0.038$ &&   $0.644\pm 0.036$
\enddata
\tablenotetext{a}{Measured with MagE spectrograph through aperture of
		  $6\farcs61\times 1\farcs0$ centered on brightest part of Nebula,
		  corresponding to a projected area of $2.12\times 0.32$ kpc$^2$ at
		  the distance of \n7252.}
\tablenotetext{b}{Ratio of fluxes corrected for Milky Way foreground extinction of
		  $\avmw = 0.10$.}
\tablenotetext{c}{Ratio of fluxes corrected for total extinction (Milky Way plus
		  internal) of $\av = 0.25 \pm 0.15$.}
\tablenotetext{d}{Reddening-corrected flux within aperture is\ \ $F_0({\rm H}\beta) =
		  (0.99\pm 0.18)\times 10^{-16}$ erg cm$^{-2}$ s$^{-1}$.}
\tablenotetext{e}{Reddening-corrected flux within aperture is\ \ $F_0({\rm H}\alpha) =
		  (3.06\pm 0.38)\times 10^{-16}$ erg cm$^{-2}$ s$^{-1}$, with
		  corresponding luminosity of\ \ $L_0({\rm H}\alpha) = (1.60\pm 0.20)
		  \times 10^{38}$ erg s$^{-1}$.}
\end{deluxetable*}

\begin{deluxetable*}{lccccccccc}
\tablecolumns{10}
\tablewidth{0pt}
\tablecaption{Comparison of Flux Ratios and Line Luminosities in the Nucleus,
	      Central Gas Disk, and \oiii\ Nebula of \n7252\label{tab04}}
\tablehead{
   \colhead{}                          &
   \colhead{}                          &
   \colhead{}                          &
   \multicolumn{3}{c}{\lineratzeromw\tablenotemark{a}} &
   \colhead{}                          &
   \multicolumn{3}{c}{\lapp\tablenotemark{b} ($10^{40}$ erg s$^{-1}$)} \\[2pt]
   \cline{4-6}  \cline{8-10}             \\[-5pt]
   \colhead{\phn Line/Ion\phn}         &
   \colhead{$\lambda$}                 &
   \colhead{}                          &
   \colhead{Nucleus\tablenotemark{c}}  &
   \colhead{Gas Disk\tablenotemark{d}} &
   \colhead{\oiii\ Nebula\tablenotemark{e}} &
   \colhead{}                          &
   \colhead{Nucleus}                   &
   \colhead{Gas Disk}                  &
   \colhead{\oiii\ Nebula\tablenotemark{f}}
}
\startdata
\oii\dotfill  & 3727 &&     $<3.7$     & $0.87\pm 0.18$ &  $7.86\pm 0.15$     &&       $<0.19$     & $4.57\pm 0.69$           & $\phn0.90\pm 0.08$\tablenotemark{g} \\
\hbet\dotfill & 4861 &&     $1.00$     &     $1.00$     &      $1.00$         &&  $0.052\pm 0.015$ & $5.25\pm 0.79$           & $\phn0.12\pm 0.01$\tablenotemark{g} \\
\oiii\dotfill & 5007 && $0.45\pm 0.21$ & $0.17\pm 0.04$ & $10.09\pm 0.25\phn$ &&  $0.023\pm 0.009$ & $0.89\pm 0.18$           & $1.17\pm 0.04$ \\
\halp\dotfill & 6563 && $3.13\pm 1.20$ & $4.83\pm 0.87$ &  $3.24\pm 0.10$     &&  $0.163\pm 0.040$ & $25.4\phn\pm 2.5\phn\phn$& $0.37\pm 0.04$
\enddata
\tablenotetext{a}{Ratio of fluxes corrected for Milky Way foreground extinction of
		  $\avmw = 0.10$.}
\tablenotetext{b}{Apparent luminosity, corrected only for Milky Way foreground (but
		  not internal) extinction.}
\tablenotetext{c}{Measured with MagE spectrograph through aperture of
		  $0\farcs90\times 0\farcs70$ ($289\times 225$ pc$^2$).}
\tablenotetext{d}{Integrated within simulated ring-shaped aperture of radius $r = 0\farcs5$\,--\,4\arcsec\ = 0.16\,--\,1.28 kpc.}
\tablenotetext{e}{Measured in brightest part of nebula with MagE (see Footnote a in Table 3).}
\tablenotetext{f}{Apparent luminosity is for entire nebula.}
\tablenotetext{g}{Estimated from \lapp(H$\alpha$) and line ratio measured in MagE slit.}
\end{deluxetable*}

\begin{deluxetable*}{lccccc}
\tablecolumns{5}
\tablewidth{0pt}
\tablecaption{Comparison of \oiii\ Nebula with Hanny's Voorwerp\label{tab05}}
\tablehead{
\colhead{}                  &
\colhead{}                  &
\colhead{}                  &
\colhead{\n7252}            &
\colhead{IC 2497}           &
\colhead{}                     \\
\colhead{\phm{WWWWW}Parameter\phm{WWWWW}} &
\colhead{Symbol}            &
\colhead{}                  &
\colhead{\oiii\ Nebula}     &
\colhead{Hanny's Voorwerp\tnm{a}} &
\colhead{Units}
}
\startdata
Galaxy:                           &                     &&                      &                            &           \\
\ \ Heliocentric velocity\dotfill & $cz_{\rm hel}$      && 4749                 &  15,056                    & \kms      \\
\ \ Distance\dotfill              & $D$                 && 66.2                 &  220                       & Mpc       \\

Emission-line nebula:             &                     &&                      &                            &           \\
\ \ Projected size\dotfill        & \nodata             && $10.1\times 6.6$     & $33\times 18$              & kpc       \\
\ \ Projected distance\tnm{b}\dotfill&  $r_{\rm p}$     && $\sim\,$5.4          & $\sim\,$24                 & kpc       \\
\ \ Velocity range\tnm{c}\dotfill & \dvlos		&& $-$50 to $-$200      & $-$220 to $-$300           & \kms      \\
\ \ \oiii\ luminosity\dotfill     & $L_0(\lambda 5007)$ && $1.37\times 10^{40}$ & $\sim 1.5\times 10^{42}$   & \ergsec   \\
\ \ \hbet\ luminosity\dotfill     & $L_0(\hbet)$        && $1.36\times 10^{39}$ & $1.4\times 10^{41}$        & \ergsec   \\
\ \ Excitation ratio\dotfill  & $L_0(\lambda 5007)/L_0(\hbet)$&& $10.0\pm 0.5$  & $10.5\pm 1.0$              &           \\
\ \ Electron density\dotfill      & \nel                && $1 \la \nel \la 300$ & $<50$                      & cm$^{-3}$ \\
\ \ Electron temperature\dotfill  & $T$                 && $\la 8900$           & $13,500\pm 1300$           & K         \\
\ \ Metallicity\dotfill           & $Z/Z_{\odot}$       && \phs$0.75\pm 0.10$   & $0.30\pm 0.15$             &           \\
\ \ Ionization parameter (log)\dotfill &  $\log U$      && $-2.8\pm 0.1$        & $-2.2$                     &           \\

Nucleus:                          &                     &&                      &                            &           \\
\ \ Past ionizing luminosity\tnm{d}\dotfill&
                                        $L_{\rm ion}$   && $\ga 5\times 10^{42}$& $\ga\,$(2--7)$\times 10^{45}$&\ergsec  \\ 
\ \ Present X-ray luminosity\dotfill& \lxttzero         && $<5\times 10^{39}$   & $\la 4.2\times 10^{40}$    & \ergsec   \\
\ \ Decline time\tnm{e}\dotfill   & $\Delta t$          && 20\,--\,200          & 40\,--\,230                & kyr
\enddata
\tnt{a} {Data taken from \citet{lint09}, \citet{scha10}, and \citet{keel12b}.}
\tnt{b} {From nucleus to brightest part of nebula.}
\tnt{c} {{L}ine-of-sight velocities relative to galaxy nucleus.}
\tnt{d} {Inferred from \hbet\ flux of \oiiineb\ (see \S~\ref{sec43}).}
\tnt{e} {Time since the most recent enhanced AGN activity, still seen by nebula, ceased.}
\end{deluxetable*}


\end{document}